\documentclass[11pt,a4paper]{article}                           

\usepackage{amsmath}
\usepackage{amssymb}
\usepackage[usenames]{color} 
\usepackage{multirow}
\usepackage{graphicx}
\usepackage{epstopdf}
\usepackage{array}
\usepackage{hhline}
\usepackage{cite}
\usepackage{microtype,colortbl}
\usepackage[bf]{caption}
\usepackage{color}
\usepackage[usenames,dvipsnames]{xcolor}
\usepackage{subcaption}
\usepackage{pstool}
\usepackage{bbold}
\usepackage{tikz}
\usepackage{ytableau}
\usepackage{braket}
\usepackage[T1]{fontenc}
\usepackage[framemethod=TikZ]{mdframed}
\usepackage{lipsum}
\mdfdefinestyle{MyFrame}{%
    linecolor=gray!80!white,
    outerlinewidth=0.5pt,
    roundcorner=10pt,
    innertopmargin=\baselineskip,
    innerbottommargin=\baselineskip,
    innerrightmargin=20pt,
    innerleftmargin=20pt,
    backgroundcolor=gray!15!white}

\usepackage{ifpdf}
\ifpdf
  \usepackage[pdftex,
    pdftitle={},
    pdfauthor={},
    pdfsubject={},
    bookmarksopen, bookmarksnumbered, bookmarksopenlevel=2]{hyperref}
\fi
\def\hybrid{\topmargin -20pt    \oddsidemargin 0pt
        \headheight 0pt \headsep 0pt
        \textwidth 6.25in       
        \textheight 9 in       
        \marginparwidth .875in
        \parskip 5pt plus 1pt 
          \jot = 1.5ex
   }
\hybrid
\numberwithin{equation}{section}
\numberwithin{table}{section}

\newcommand{\beq}{\begin{equation}}  \newcommand{\eeq}{\end{equation}}
\newcommand{\bal}{\begin{aligned}}   \newcommand{\eal}{\end{aligned}}
\newcommand{\bea}{\begin{eqnarray}}  \newcommand{\eea}{\end{eqnarray}}

\newcommand{\bmat}{\left(\begin{array}}
\newcommand{\emat}{\end{array}\right)}

\newcommand{\bbC}{\mathbb{C}}
\newcommand{\bbR}{\mathbb{R}}

\newcommand{\cP}{\mathcal{P}}

\newcommand{\cD}{\mathcal{D}}
\newcommand{\cL}{\mathcal{L}}
\newcommand{\cS}{\mathcal{S}}

\newcommand{\cN}{\mathcal{N}}

\newcommand{\cH}{\mathcal{H}}

\newcommand{\cM}{\mathcal M}

\newcommand{\I}{\text{Im}}

\usepackage{cancel}

\newcommand{\be}{\begin{equation}}
\newcommand{\ee}{\end{equation}}

\newcommand{\bbZ}{\mathbb{Z}}

\definecolor{Gray}{gray}{0.95}

\begin{document}

\baselineskip=15pt
\parskip 5pt plus 1pt

\vspace*{-1.5cm}
\begin{flushright}    
  {\small 
  
  }
\end{flushright}

\vspace{2cm}
\begin{center}        
{\huge Taming the Landscape of Effective Theories }

\end{center}

\vspace{0.5cm}
\begin{center}        
{\large  Thomas W.~Grimm  }
\end{center}

\vspace{0.15cm}
\begin{center}        
\emph{
Institute for Theoretical Physics \\
Utrecht University, Princetonplein 5, 3584 CE Utrecht, The Netherlands}
 
\end{center}

\vspace{2cm}


\begin{abstract}
\noindent
\baselineskip=15pt
We introduce a generalized notion of finiteness that provides a structural principle for 
the set of effective theories that can 
be consistently coupled to quantum gravity. More concretely, we propose a Tameness Conjecture 
that states that all valid effective theories are 
labelled by a definable parameter space and must have scalar field spaces and coupling functions 
that are definable using the tame geometry 
built from an o-minimal structure. We give a brief introduction to tame geometry and describe how 
it restricts sets, manifolds, and functions. We then collect evidence for the Tameness 
Conjecture by studying various effective theories arising from string theory compactifications by using some 
recent advances in tame geometry. In particular, we will exploit 
the fact that coset spaces and period mappings are definable in an o-minimal structure and  
argue for non-trivial tameness results in higher-supersymmetric theories and 
in Calabi-Yau compactifications. 
As strongest evidence for the Tameness Conjecture over 
a discrete parameter space, we then discuss a recent theorem stating that 
the locus of self-dual flux vacua of F-theory admits a tame geometry even if one allows for any flux choice 
satisfying the tadpole constraint. This result implies the finiteness of self-dual flux vacua in F-theory.

\end{abstract}

\thispagestyle{empty}

\setcounter{page}{1}


\newpage

\tableofcontents

\newpage

\section{Introduction and a conjecture}
\label{sec:intro}

In recent years the search for general principles restricting the form of 
any effective theories that can be consistently coupled to quantum gravity has attracted 
much attention \cite{Palti:2019pca,vanBeest:2021lhn}. These principles have been formulated in a number of quantum gravity 
or `swampland' conjectures. They aim to discriminate consistent effective theories that are part of the 
landscape, from those that are fundamentally flawed and reside in the swampland. 
One of such principles is the claim that the number of effective theories that are valid below a fixed 
cut-off scale that are consistent with quantum gravity is finite \cite{Douglas:2003um,Acharya:2006zw,Hamada:2021yxy}. 
The finiteness of 
effective theories implies constraints on  
the allowed scalar potentials and the scalar field spaces on which the 
effective theory is valid, since a new effective theory can arise when 
lowering the energy scale and settling in a new vacuum. Despite the fact that 
this clearly restricts valid effective theories, it has not been clear how to turn this 
into a structural principle. The aim of this work is to introduce 
a mathematical structure --a tame geometry-- and 
argue that it provides a concrete way to implement finiteness 
constraints on the set of 
consistent effective theories. Furthermore, we conjecture 
that it should be used as a novel general principle to constrain 
field spaces and coupling functions of UV-completable effective theories.

In this paper we will give a novel perspective on the set of consistent effective theories by claiming that 
the landscape admits a certain well-defined geometric structure. More precisely, we will propose a Tameness Conjecture 
that constrains the set of all effective theories that are valid up to some fixed finite energy cut-off scale and can be consistently 
coupled to quantum gravity. We conjecture that all such theories are labelled by a parameter space 
that is definable in a so-called o-minimal structure. Furthermore, we claim that also the scalar field spaces 
and coupling functions, which might depend on these parameters, are definable in the same o-minimal structure.
O-minimal structures implement finiteness on a fundamental level and are the prime example of a topologie mod\'er\'ee, a tame topology, 
envisioned by Grothendieck \cite{Grothendieck}.\footnote{Grothendieck's original motivation for introducing a new form of topology stems from the study of 
moduli spaces of Riemann surfaces and maps between them.} Grothendieck's vision was to develop 
a topology for geometers that excludes pathological situations that can arise in classical topology.
Notably, o-minimal structures can be defined over the real numbers and provide an 
extension of real algebraic geometry while keeping some of its powerful results 
intact. They thus provide us with a framework to leave the world of complex geometry, which is 
often only occurring in effective theories due to the presence of supersymmetry, while setting a completely 
new focus on finiteness and tameness. 
 It is interesting to highlight that the original interest on o-minimal structures arose 
from model theory, which is a part of mathematical logic that studies the relationship between formal theories and their models. 
By now, however, these structures have found applications in several fields of mathematics reaching from 
number theory to geometry. 

The basic strategy in defining a tame topology based on o-minimal structures \cite{vdD} is to specify the space 
of allowed subsets of $\bbR^n$, for every $n$. On this space of `tame sets', also called definable sets, one 
can then define `tame functions', which are termed definable functions. Hereby one always means that these 
sets and functions are defined with respect to a specified o-minimal structure.
The fundamental tameness property of each o-minimal structure is the fact that the only 
definable sets in one real dimension are the finite union of points and intervals.  
This property becomes powerful when combined with the requirement that all linear 
projections of higher-dimensional sets eventually reduce to sets of this type on the real line.
Tame topology hereby treats a connected set of infinitely many points, such as 
an interval or the full real line, as a single object. While the 
simplest example of an o-minimal structure is formed by collecting  
sets that are defined by polynomial equalities and inequalities, the 
existence of much richer o-minimal structures will be central in this work. Firstly,
it is a remarkable mathematical fact that extensions exist in which the  
sets can be defined by also using transcendental functions. In particular, an important 
result of Wilkie \cite{Wilkie96}, which states that adding the real exponential function does not violate the 
tameness axioms, has allowed mathematicians to use o-minimal structures in 
a wide set of geometric applications, such as the Hodge theory application that we will exploit 
in this work. Secondly, it is apparent that such extensions are needed to describe well-known physical 
settings since many effective theories cannot be described by purely algebraic data. 
In particular, instanton corrections to coupling functions should be consistent with 
the Tameness Conjecture and hence describable within tame geometry. It is interesting 
that o-minimal structures are, on the one hand, rich enough for many applications, while on 
the other hand they possess strong finiteness constraints.\footnote{Note that in \cite{Cecotti:2021cvv} a geometric framework, which was called `domestic geometry', was introduced
to describe certain UV-completable effective theories. 
It would be interesting to investigate the relation of this proposal to the tame geometry used here.}

To provide evidence for the Tameness Conjecture, we will have a detailed look at some of the 
well-understood effective actions derived from string theory. String theory has only one free parameter, the string length, and the ten-dimensional effective supergravity 
actions admit simple scalar field spaces for the arising massless scalars. At the two-derivative level the 
Tameness Conjecture is readily shown in these highly supersymmetric settings. 
However, if we consider the theories on a compact manifold, 
it is well-known that a plethora of effective theories will arise in less than ten dimensions. 
The scalar field spaces of these effective theories can be very involved and numerous new 
parameters arise from the geometry of the compactification space and possible backgrounds 
for the other fields of the higher-dimensional theory, such as background fluxes. 
It turns out that 
in theories with more than 8 supercharges, supersymmetry together with some simple-to-state 
finiteness conditions already ensures that the Tameness Conjecture holds. We will argue that this conclusion requires us to 
use some recent mathematical results about the tameness of double cosets, 
i.e.~arithmetic quotients of the form $\cM_{\Gamma,G,K} = \Gamma \backslash G / K$. 
The Tameness Conjecture is then satisfied if the free parameter choices, e.g.~labelling the allowed groups $G$ and $\Gamma$, are finite. 
Showing finiteness statements of this type is the aim of much current research \cite{Kumar:2009us,Kumar:2009ae,Kumar:2010ru,Adams:2010zy,Park:2011wv,Kim:2019vuc,Lee:2019skh,Kim:2019ths,Dierigl:2020lai,Font:2020rsk,Font:2021uyw,Hamada:2021bbz,Tarazi:2021duw}. 

When reducing the amount of supersymmetry, the Tameness Conjecture provides a more independent criterium from this symmetry, 
since one can find field spaces and coupling functions that are compatible with supergravity but are not tame. 
Nevertheless, we will show that in some of the best understood string compactifications we only 
encounter field spaces and coupling functions that are definable in an o-minimal structure. More precisely, we will 
look at compactifications of Type II string theory on Calabi-Yau threefolds leading to four-dimensional effective theories 
with $\cN=2$ supersymmetry. In these cases the field spaces are built from the moduli spaces of the compact geometry 
and we will argue that these admit a tame geometry. Moreover, we will introduce a recent foundational result 
of Bakker, Klingler, Tsimerman \cite{BKT} that shows that the period mapping is definable in an o-minimal structure denoted 
by $\bbR_{\rm an,exp}$. We use this result to argue that at least in the vector sector 
of $\cN=2$ actions arising from Calabi-Yau compactifications the scalar field space metric and gauge coupling function are definable.  
This provides a very non-trivial test of the Tameness Conjecture if one makes a choice for the topology of the Calabi-Yau manifold.  
Picking different topologies should be viewed as picking different discrete parameters of the effective theory and the Tameness Conjecture 
asserts that there are only finitely many such choices. 

It is a central statement of the Tameness Conjecture that all viable scalar potentials are 
definable in an o-minimal structure. This statement ensures finiteness when lowering the cut-off scale 
of the theory further. Indeed, if after lowering the cut-off some of the fields are too heavy and need 
to be integrated out, tameness of the original scalar potential will ensure that the resulting new 
low-energy scalar field space is also definable in an o-minimal structure. 
In the last part of this work we will provide evidence for this property of the scalar potential 
in flux compactifications of Type IIB string theory and F-theory reviewed in \cite{Grana:2005jc,Douglas:2006es,Denef:2008wq}. 
These compactifications yield a well-understood class of effective theories with $\cN=1$ supersymmetry that 
admit a positive definite scalar potential solely induced by 
background fluxes. The Minkowski vacua of this potential arise if the fields adjust such that the fluxes become  
(imaginary) self-dual. These vacua admit well-defined lifts to higher dimensions and we expect the effective theory 
with $\cN=1$ or $\cN=0$ obtained when integrating out the massive scalar field to be well-behaved. 
We will argue that the flux scalar potential is definable in the o-minimal structure $\bbR_{\rm an, exp}$
following \cite{BKT} for fixed fluxes. In this setting, however, we can go further and treat the fluxes as
discrete parameters. Definability is retained if the fluxes satisfy the tadpole cancellation condition and we 
consider the potential sufficiently close to a Minkowski vacuum. 
This will follow from a result of Bakker, Schnell, Tsimerman, and the author \cite{BGST}, which states that  
the locus of self-dual fluxes is definable in the o-minimal structure $\bbR_{\rm an, exp}$. We will briefly 
summarize the argument and explain how it shows the finiteness of flux choices. Evidence for such 
a finiteness result has appeared previously in \cite{Douglas:2004zu,Douglas:2004kc,Douglas:2005df,Douglas:2006zj,Lu:2009aw,Grimm:2019ixq,Grimm:2020cda}. 
The theorem of \cite{BGST} generalizes 
a famous theorem of Cattani, Deligne, Kaplan \cite{CDK} proving the finiteness of Hodge classes satisfying 
a `tadpole cancellation condition'.

Let us close by stating the Tameness Conjecture in a weak and a strong form, where the latter 
specifies an o-minimal structure that suffices in all considered string theory examples:
\begin{mdframed}[style=MyFrame]
\textbf{Tameness Conjecture}\\[.1cm]
 All effective theories valid below a fixed finite energy cut-off scale that can be consistently coupled to quantum gravity are labelled by a definable parameter space and must have scalar field spaces and coupling functions that are definable in an o-minimal structure.
  
\noindent
\textbf{Strong Tameness Conjecture}\\[.1cm] 
The o-minimal structure that makes the effective theory definable is $\mathbb{R}_{\rm an,exp}$.
\end{mdframed}

This paper is organized as follows. In section \ref{sec:effective_theories} we explain in more 
detail which aspects of an effective theory we are considering in this work. In particular, we introduce the
relevant notion of parameter space, scalar field space, the coupling functions of an effective theory. We then 
comment on various effective theories arising in string compactifications and highlight additional challenges 
that need to be faced when a scalar potential is present. In section \ref{intro-to-o} we then give a lightning introduction to 
o-minimal structures and tame topology with a focus on some of the foundational results. This will 
help to clarify the statement of the Tameness Conjecture and provide the background for the more 
advanced results used in the third part of this work. In fact, in section \ref{tamegeom-landscape}  
we will introduce the evidence for the Tameness Conjecture, by discussing various string 
theory compactifications. In particular, we will also sketch the argument that the flux scalar 
potential is a tame function and that there are only finitely many self-dual fluxes. 

\section{On effective theories and their coupling functions} \label{sec:effective_theories}

The Tameness Conjecture claims that the scalar field spaces and coupling functions 
in any effective theory that can be consistently coupled to quantum gravity 
are definable in a tame geometry introduced in section \ref{intro-to-o}. To make this more concrete 
let us consider a set of scalar fields  $\phi^i$ and gauge fields 
$A^C$ coupled to Einstein gravity. In addition to these fields, the effective 
theory can also contain other fields, such as fermions or higher-form fields, but we will not display them in 
the following. Then the Lagrangian of the effective theory then schematically takes the form 
\beq \label{general_action}
   \cL = R -  g_{ij}(\phi,\lambda)\ D_\mu \phi^i D^\mu \phi^j  - f_{AB}(\phi,\lambda)\ F^A_{\mu \nu} F^{B\, \mu \nu} - V(\phi,\lambda) + \ldots \ ,
\eeq
where $V$ is the scalar potential of the theory. 
Let us denote by $\cM_\lambda$ the field space spanned by the $\phi^i$ with metric $g_{ij}$.
In general, the coupling functions 
$g_{ij},  f_{AB},V,\ldots$ will vary over $\cM_\lambda$. In addition, we will allow for 
the field space $\cM_\lambda$ and the coupling functions to depend on a set of parameters $\lambda_\alpha$, 
which we consider to be part of a 
parameter space $\cP$. These parameters 
can be vacuum expectation values of fields that have been integrated out, or they can be discrete 
parameters. Hence, the space $\cP$ does not have to be a smooth manifold, but rather can be 
just some general set. The Tameness Conjecture both restricts the geometry of the set 
\beq \label{def-domain}
    \mathcal{D} = \{ (\phi^i , \lambda_\alpha), \ \phi \in \cM_\lambda,\, \lambda \in \cP \} 
\eeq
as well as the behavior of the coupling functions, such as $g_{ij}, f_{AB}, V$. The crucial point is here, that we view 
these coupling functions as maps valued on $ \mathcal{D} $ with a set of special 
tameness properties introduced in section \ref{intro-to-o}. 

In this section we recall some effective theories  arising in compactifications of string theory.
This will allow us to highlight some necessary requirements 
on the geometry of the Tameness Conjecture that need to be satisfied in order that it is general enough to apply to well-understood 
examples. 
Clearly, our discussion will not be exhaustive and should only be seen as a motivation for the 
structures introduced later. In a first step, we will concentrate on theories without scalar potential in subsection \ref{sec:Scalar-coupling}. 
We discuss the inclusion of a scalar potential in subsection \ref{sec:scalar_pot} and point out some additional complications 
arising in this case. The reader familiar with string compactifications does not need to spend much time on this section.

\subsection{On scalar field spaces and coupling functions in string compactifications} \label{sec:Scalar-coupling}

String theory is originally formulated in ten space-time dimensions. We note that already in ten dimensions 
all five string theories have massless scalar fields. In particular, Type IIB string theory has 
a complex scalar $\tau$, the dilaton-axion that takes values on a field space $\text{SL}(2,\bbZ)\backslash SL(2, \bbR)/SO(2)$. 
This space is non-compact, but admits a complex algebraic structure. 
While this space has much structure, it turns out that this is not a general feature of 
the field spaces arising in string theory, but rather a remnant of supersymmetry. 
In particular, the complex algebraic structure is not 
necessarily present when looking at string compactifications. 
To see this, recall that the moduli space of a torus $T^d$ is the arithmetic quotient, sometimes called 
double coset, $SO(d,d;\bbZ)\backslash SO(d,d;\bbR)/ SO(d)\times SO(d)$. Purely for dimensional reasons 
this space is not always complex. The fact that such arithmetic quotients arise as field spaces 
can be tied to the presence of some supersymmetry in the effective theory. In fact, for more than 
8 supercharges, the field spaces take the general form 
\beq \label{moduli-space-quotient}
   \cM_{\Gamma,G,K}  = \Gamma \backslash G / K \ ,
\eeq
where $\Gamma$ is a lattice and $K$ is a maximal compact subgroup of $G$.\footnote{For later purposes, we will require that $G = G(\bbR)^+$ is 
the real Lie group connected component of the identity of $G(\bbR)$, where $G(\bbR)$ is the real version of a connected linear semi-simple 
algebraic $\mathbb{Q}$-group $G(\mathbb{Q})$. The discrete group $\Gamma \subset G(\mathbb{Q})^+$ is assumed to be a torsion-free arithmetic lattice.} 
In these supersymmetric theories also the coupling functions take 
a particularly simple form. Roughly speaking these functions can 
always expressed as (quotients of) polynomials in a suitable set 
of coordinates on the field space $\cM$. This simple form is compatible with 
the fact that instanton corrections are often forbidden by supersymmetry.

More involved examples of field spaces and coupling functions arise when compactifying string theory 
on a Calabi-Yau threefold such that the 
resulting four-dimensional theory has 8 supercharges or less. 
If one insists that the Calabi-Yau condition is preserved 
then the deformation spaces of these spaces split into complex structure 
and K\"ahler structure deformations. In the following we 
will review some facts about the complex structure moduli space $\mathcal{M}_{\rm cs}$, keeping in mind that 
the geometry of the K\"ahler  structure moduli space is a special case of this more general discussion 
after using mirror symmetry. For polarized Calabi-Yau threefolds $Y_3$
the moduli space $\mathcal{M}_{\rm cs}$ is quasi-projective \cite{Viehweg} and non-compact. It has complex dimension 
$h^{2,1}= \text{dim} H^{2,1}(Y_3)$ and we will use local coordinates $z^i$, $i=1,..., h^{2,1}$ in the following.
The natural metric on $\mathcal{M}_{\rm cs}$ that arises in string theory effective actions
is the so-called Weil-Petersson metric $g_{i\bar \jmath}$. This metric is K\"ahler and can be derived from 
a K\"ahler potential $
   K = - \log i \bar \Pi^{I} \eta_{IJ} \Pi^J $. 
Here $\eta_{IJ} = \gamma_I \cap \gamma_J$ is the intersection matrix of a basis of three-cycles $\gamma_I $ and
we have abbreviated 
\beq \label{periods}
   \Pi^I (z)= \int_{\gamma_I} \Omega\ .
\eeq
These integrals are known as period integrals, or periods for short, of the, up to rescaling, unique $(3,0)$-form $\Omega$. 
The resulting metric takes the form 
\beq \label{N=2metric}
    g_{i\bar \jmath} = \partial_{z^i} \partial_{\bar z^j} K = \frac{\eta_{IJ} (D_i \Pi^I) (\overline{D_{j} \Pi}\phantom{\Pi}^{\!\!\! \! \! \! J})}{\eta_{KL} \Pi^K \bar \Pi^L }\ ,
\eeq
where $D_i \Pi^I = (\partial_{z^i} + \partial_{z^i} K ) \Pi^I$. 

The periods also determine some of the other couplings of the effective theory. 
For example, consider Type IIB string theory on $Y_3$. In the four-dimensional effective theory arising after compactification 
also the gauge coupling functions $f_{A B}$ for the R-R U(1)s can be expressed in terms of the periods $\Pi$.
To explicitly give $f_{AB}$, we first need to introduce a symplectic homology basis $\gamma_I = (\gamma^A, \tilde \gamma_B$, 
such that $\gamma^A \cap  \gamma^B = \tilde \gamma_A \cap \tilde \gamma_B = 0$ and $\gamma^A \cap \tilde \gamma_B = \delta^A_B$. This allows us to split 
$\Pi = (\Pi^A,\Pi_B)$ and the $\cN=2$ gauge coupling function is then given by 
\beq \label{N=2gaugecoupling}
    f_{AB} = (\Pi_A , \overline{D_{j} \Pi}\phantom{\Pi}_{\!\!\! \! \!A}) (\Pi^B, \overline{D_j \Pi}\phantom{\Pi}^{\!\!\! \! \!B})^{-1}\ . 
\eeq
Hence, in order 
that the Tameness Conjecture for coupling functions can possibly be true, it has to hold at least for the couplings \eqref{N=2metric} and \eqref{N=2gaugecoupling} derived 
from the period map.

The 
periods $\Pi^I$ are holomorphic but, in general, complicated transcendental functions. However, it is known from the 
work of Schmid \cite{Schmid} that in a sufficiently small neighbourhood near every boundary of $\mathcal{M}_{\rm cs}$ they admit an expansion that 
splits them into a polynomial plus exponentially suppressed part. Let us pick coordinates $t^\alpha,\zeta^a$, such that 
the considered boundary is at $t^\alpha = i \infty$ and $\zeta^a$ finite. Then one can expand \footnote{Note that this  requires us 
to work on the universal cover of the local boundary neighbourhood. Furthermore, we allow so-called base changes $t^\alpha \rightarrow (t^\alpha)^n$ 
to reach the general form \eqref{nilp-orbit}.} 
\beq \label{nilp-orbit}
    \Pi = e^{t^i N_i }\Big( a_0 + \sum_{r_i} e^{2\pi r_i t^i}a_{r_1,...,r_n}\Big)\ , 
\eeq
where the $N_i$ are nilpotent matrices and the coefficients $a_{\bullet}$ can still vary holomorphically with $\zeta^a$. 
Focusing on the behaviour in the $t^i$, this implies that the metric $g_{ij}$ will in general involve  
finitely many polynomial terms as well as a host of exponentially suppressed corrections. This rather constrained 
behaviour in the asymptotic, non-compact directions, will reappear in a much more general way in the 
tame geometry introduced in section \ref{intro-to-o}. In fact, it will turn out to be one of the hallmarks of tameness that only a certain set of functions 
can arise on such non-compact tails. Before explaining this in detail, let us discuss some further issues that arise when one 
includes a scalar potential.

\subsection{Scalar potentials and the challenges to implement finiteness} \label{sec:scalar_pot}

An additional challenge in understanding the structure of the landscape of the effective theories 
arises when one includes a potential for the fields, since then a cut-off dependence is apparent. 
To make this clearer, let us consider an effective theory with a cut-off $\Lambda$. For simplicity, we will only discuss bosonic 
scalars $\phi^i$ in the following and focus on the scalar potential $V(\phi,\lambda)$. The scalar potentials 
varies over $\cD$ defined in \eqref{def-domain}, where $\cM_\lambda$ is the field space and $\cP$ is a space of parameters. 
The notion of effective potential and $\cM_\lambda$ will change when lowering the cut-off, say to $\hat \Lambda < \Lambda$.
In this case some of the $\phi^i$ might have masses above this scale and have 
 to be integrated out. Classically, this can be done by solving the vacuum conditions $\partial_{\phi^k} V=0$
 for the massive fields. 
 Clearly, there might be several solutions to this equation and, depending 
 on our choice of solutions, we end up with a different effective theory. The field space $\cM(\Lambda)$
 can thus reduce to $\cM(\hat \Lambda) = \cup_\alpha \cM_{\alpha}$, where $\cM_{\alpha}$ is the field space associated to 
 the $\alpha$th effective theory. To each of these theories a parameter space $\cP_\alpha$ can 
 appear, which now might include the vacuum expectation values of the fields that have been integrated out. 
 Note that if we continue lowering the cut-off, eventually only the massless 
 fields with a moduli space will remain and the effective theories will not have any potential.
 
The Tameness Conjecture claims that there is a constraint on allowed scalar potentials. It was 
motivated by the aim to implement finiteness of effective theories below a certain cut-off. 
Hence, we can again highlight some of the necessary properties of the tame geometry such 
that this is actually achieved. 
In fact, finiteness is to demand that for every viable $\cM$ and $V(\phi)$ only finitely many $\cM_{\alpha}$ can arise when 
lowering the cut-off. In particular, this implies that the scalar potential 
has only finitely many minima. It is 
easy to think of functions that violate such a condition. Clearly, some periodic function 
such as sin$(\phi)$ has infinitely many vacua distributed over the real line, but we can also accumulate vacua near $\phi=0$,
by considering  
\beq \label{counter}
  V(\phi) = \text{sin}(\phi^{-1})\ , \qquad  V(\phi) = \phi^8 \text{sin}(\phi^{-1})\ .
\eeq
As discussed in \cite{Acharya:2006zw}, these functions appear to be not very special and, at first, it seems very hard to state a principle that excludes 
these choices as viable scalar potentials.\footnote{A physical proposal was made in \cite{Acharya:2006zw}, where 
it was suggested that one should revise the notion of vacuum taking into account tunneling and heights 
of barriers between vacua.} However, the tame geometry introduced in section \ref{intro-to-o} 
actually gives precisely such a restriction. This then implies that potentials of the form \eqref{counter}
should not appear in string compactifications. 

One of the best understood string theory compactifications that leads to a four-dimensional 
effective theory with minimal supersymmetry are Type IIB orientifold compactifications 
with O3/O7-planes and a flux background \cite{GKP,Grimm:2004uq,Grana:2005jc,Douglas:2006es}. The compactification space 
is, up to a conformal factor, a Calabi-Yau threefold $Y_3$ supplemented by an orientifold 
involution. Before including background fluxes the complex structure 
deformations of $Y_3$ that are compatible with the orientifold involution are flat directions of 
the effective theory, i.e.~they do not receive a mass through a classical potential. This changes 
when including background fluxes $H_3,F_3 \in H^{3}_-(Y_3,\bbZ)$, which are non-trivial background 
values of the field-strengths of the NS-NS and R-R two-forms of Type IIB string theory that are compatible with 
the orientifold involution. These fluxes 
are constrained by a tadpole cancellation condition
   $\int_{Y_3} F_3 \wedge H_3 = N_{\rm b}$, 
where $N_{\rm b}$ can be derived when studying the background source terms for D-branes and O-planes 
that are included in the setting. 
$N_{\rm b}$ is a fixed integer number independent of the fluxes.
The resulting 
scalar potential can then be given in terms of $G_3 = F_3-\tau H_3$, where $\tau$ is the dilaton-axion 
of Type IIB string theory. It takes the form
\beq
   V(z,\tau,G_3) = \frac{c}{8\, \I \tau} \int_{Y_3} (\bar G_3 - i \star \bar  G_3) \wedge \star (  G_3+ i \star   G_3) \ ,
\eeq
where the coefficient $c$ can depend on the volume of $Y_3$, which will be irrelevant in this section. 
This scalar potential non-trivially depends on $\tau$ and the complex structure deformation in $\cM_{\rm cs}$ 
compatible with the orientifold involution. We have indicated this dependence by introducing complex scalar fields $z^i$ as local coordinates 
on $\cM_{\rm cs}$. Compared with our general discussion after \eqref{general_action}, we thus have a field space $\cM$ containing 
$\text{SL}(2,\bbZ)\backslash SL(2, \bbR)/SO(2) \times \cM_{\rm cs}$
and a parameter space containing the flux lattice $H^{3}_-(Y_3,\bbZ) \times H^{3}_-(Y_3,\bbZ)$. 

We now note that the scalar potential can be written as a norm-square of the complex 
flux $iG_3- \star  G_3$ when introducing the Hodge norm 
\beq \label{Hodge-positive}
    \|\omega \|^2 = \int_{Y_3} \bar \omega \wedge \star \omega \ > 0 \ ,
\eeq
which is non-vanishing for a non-trivial element $\omega \in H^3 (Y_3,\bbC)$. Hence, we find that 
$V(z,\tau,G_3) \geq 0$ and global minima of this potential are obtained when \footnote{Note that the condition $\star G_3 =   iG_3$ should be read as condition in the cohomology group $H^{3}(Y_3,\bbC)$. }
\beq\label{im_self-dual}
  V(z,\tau,G_3) = 0 \qquad \Leftrightarrow \qquad  \star G_3 =   iG_3\ .
\eeq
Recalling the aim of establishing finiteness properties, we might thus 
ask if the number of `distinct' solutions of \eqref{im_self-dual} is finite if one is allowed to also chose 
the fluxes $H_3,F_3$ that satisfy the tadpole bound. Here we count different flux choices and different 
connected components $\cM_\alpha \subset \cM_{\rm cs}$, which means that there could be flat directions 
that are not stabilized by \eqref{im_self-dual}. As we will explain below it is, in fact, true that the number of 
solutions is finite. In other words, the Hodge star as a function on complex structure  moduli space $\cM_{\rm cs}$
must be special and, in particular, potentials that are similar to the ones appearing in \eqref{counter} should not occur.

Let us note that one might wonder if the restriction to a weakly coupled orientifold setting is relevant for finiteness. 
In the above expressions one actually has to assume $\I \tau \gg 1$, since the string coupling $g_s$
is related to the vacuum expectation value $\langle \I \tau \rangle = g_s^{-1}$. In order to extend to 
all values of $\tau$ it is best to realize the orientifold setting directly in F-theory. The compactification 
geometry in this case is an elliptically fibered Calabi-Yau fourfold $Y_4$ and $\tau$ becomes part of the 
complex structure moduli space of this higher-dimensional geometry. Furthermore, the fluxes $F_3,H_3$ 
lift to a single four-form flux $G_4$. The scalar potential in this case takes the form \footnote{We recall here that the computation of this potential 
is done via M-theory and a subsequent lift to F-theory \cite{Denef:2008wq,Grimm:2010ks}. The M-theory solution and three-dimensional effective action has been studied 
in detail in \cite{Grimm:2014efa,Grimm:2015mua}, which gives much confidence in this setting.} 
\beq \label{potential_F}
   V = c \|G_4 - \star G_4 \|^2\ ,
\eeq
where $c$ is independent of $G_4$ and the complex structure moduli of $Y_4$.
We again look at the global minima of this potential and note that $G_4$ is constrained by a tadpole cancellation 
condition. To focus on the complex structure moduli dependence of \eqref{potential_F}, we impose the primitivity condition
 $J\wedge G_4=0$, where $J$ is the K\"ahler form. Restricting to such primitive $G_4 \in H^{4}_{\rm prim}(Y_4,\bbR)$, we then 
 look in F-theory at the solutions of the conditions  
\beq \label{G4-self-dual}
  G_4 \in H^{4}(Y_4,\bbZ)\ , \qquad \star G_4 = G_4\ , \qquad \int_{Y_4} G_4 \wedge G_4 = N'_{\rm b}\ ,
\eeq
where $N_{\rm b}'$ is again a fixed integer number. The finiteness claim now concerns the solutions to \eqref{G4-self-dual} 
and states that this equation is solved only along finitely many connected components $\cM_\alpha$ in the complex structure moduli 
space of the fourfold $Y_4$ together with finitely many 
different fluxes $G_4^\alpha$. Formulated as a condition on $V(z,G_4)$, we would like to check that this potential has only finitely 
many zero-loci. As in the orientifold setting this implies that $V$ is a special function. In the next section we will 
introduce the mathematical framework that allows us to make this more precise.

\section{A brief introduction to tame geometry} \label{intro-to-o}
 
In this section we give a lightning introduction to the theory of  o-minimal structures that define a 
form of  tame topology of $\bbR^n$. This topology is more constrained, but can nevertheless be used to 
introduce manifolds, morphisms, and many other objects familiar when defined using `ordinary' topology 
of $\bbR^n$. The resulting tame geometry is the base of the Tameness Conjecture  and 
implements a general notion of finiteness. An introduction to the basics of tame topology and o-minimal structures 
is the book by van den Dries \cite{vdD}. For a shorter summary including also some of the recent 
results connecting tame geometry with Hodge theory the reader may consult the lecture notes 
of Bakker \cite{Bakkernotes} or the brief discussion in \cite{BGST}.

\subsection{O-minimal structures and definable sets, functions, and manifolds} \label{sec:definablesets+maps}

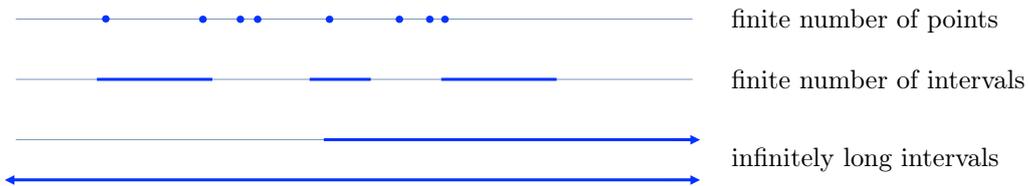
\begin{figure}[h!]
\begin{center}
 \includegraphics[width=0.6\textwidth]{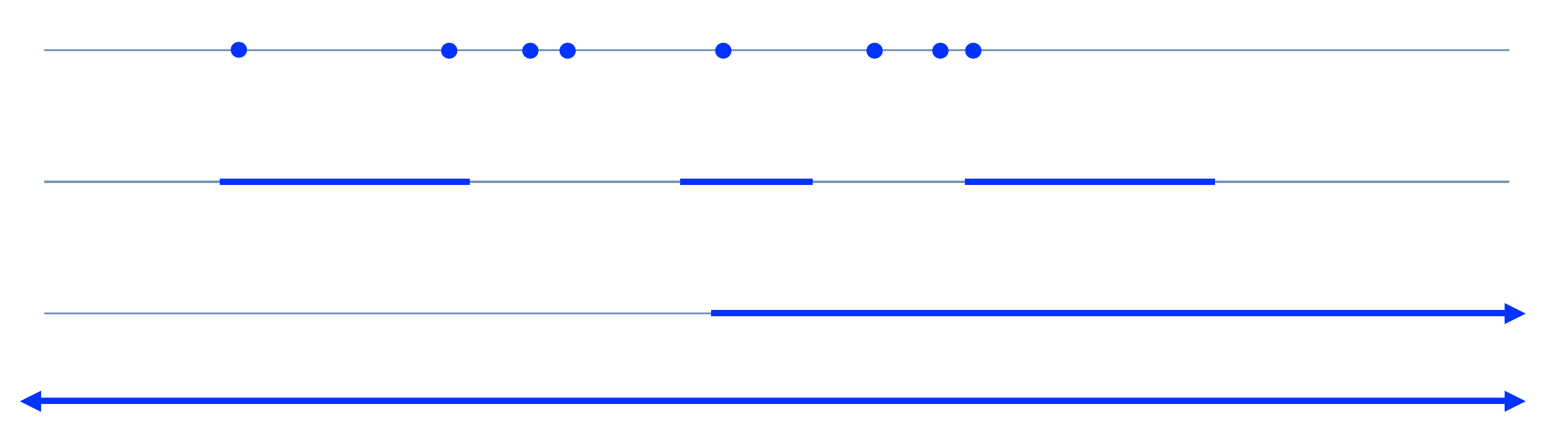} \hspace*{4.5cm}
\caption{Examples of definable sets of $\bbR$. \label{definablesetsonR}}
\end{center}
\begin{picture}(0,0)
\put(300,110){\small finite number of points}
\put(300,87){\small finite number of intervals}
\put(300,58){\small infinitely long intervals}
\end{picture}
\vspace*{-.6cm}
\end{figure}

The rough idea behind the definition of an o-minimal structure $\cS$ is the following. It will contain subsets
of all $\bbR^n$, $n=1,2,..$, which will be called $\cS$-definable, or definable for short, that give an intermediate notion between sets of  
solutions to finitely many real algebraic 
equations and the general set of subsets of $\mathbb{R}^n$. One demands that the sets contain 
any finite union, finite intersection, complements, and 
 Cartesian product of other $\cS$-definable sets. Crucially, we also require that any linear projection of a definable set is still a definable set. With this 
requirement at hand, we can implement a \textit{finiteness constraint}, i.e.~ensure the tameness of the structure, by demanding that any projection
 to the real line always yields an union of finitely many points or intervals. The latter can be closed or open and even infinitely long, see figure \ref{definablesetsonR}. 

\noindent
\textbf{O-minimal structures and definable sets:} Let us give the full definition of an o-minimal structure. 
An o-minimal structure $\cS$ is given by a collection $\cS_n$ of subsets of $\bbR^n$
with $n\geq 1$ with the following properties 
\begin{itemize}
 \item[1.] The zero-set of any polynomial $P$ in $n$ variables is in $\cS_n$;
\vspace{-.1cm}
\item[2.] each $\cS_n$ is closed under finite intersections, finite unions, and complements; 
\vspace{-.1cm}
\item[3.] if $A \in \cS_n$ and $B \in \cS_m$, then $A \times B \in \cS_{n+m}$;  
\vspace{-.1cm}
\item[4.] if $\pi: \bbR^{n+1} \rightarrow \bbR^n$ is a linear projection and $A \in \cS_{n+1}$, then $\pi(A) \in \cS_n$;
\vspace{-.1cm}
\item[5.] the set $\cS_1$ consists of finite unions of points and intervals. 
\end{itemize}
The elements of $\cS_n$ are called the $\cS$-definable sets of $\bbR^n$.

\noindent
\textbf{Definable maps:} Having introduced the notion of an o-minimal structure, we can now define what we mean by a tame map in this setting. A map $f:A\rightarrow B$
between two $\cS$-definable sets is called a $\cS$-definable map if its graph is an $\cS$-definable subset of $A\times B$. The notion of definable maps will 
be central in the following. For simplicity we will often drop the $\cS$ and call the sets and maps to be definable. 
Some basic results for definable maps are:  (1) the image and preimage of a definable set under a definable map is definable; (2) 
the composition of two definable maps is definable. 

\noindent
\textbf{Definable topological spaces and manifolds:}
Given these definitions we can now proceed by defining an  $\cS$-definable topological space  $M$. In order to do that one first introduces a 
definable atlas as a finite open covering $\{U_i \}$ of $M$ and a set of homeomorphisms $\phi_i: U_i \rightarrow V_i \subset \bbR^{n_i}$.
Definability is imposed by requiring that (1) the $V_i$ and the pairwise intersections $\phi_i(U_i \cap U_j)$ are definable sets, and 
(2) the transition functions are definable. Such a definable topological space can be upgraded to a  $\cS$-definable manifold        $\cM$
by requiring that the $V_i$ are open subsets of $\bbR^n$ and the transition functions are smooth. Complex definable 
manifolds are then obtained by viewing $\bbC^n \cong \bbR^{2n}$ and requiring the transition functions to be holomorphic.
We can now develop this further and introduce definable subsets, definable morphisms, definable analytic spaces etc. The reader can consult \cite{vdD} for further details. More important 
for the purpose of this work is to highlight in subsection \ref{sec:definable-functions+cells} some of the implications that follow from imposing $\cS$-definability. 
Before doing that it is crucial to give actual examples of o-minimal structures.

\noindent \textbf{Examples of o-minimal structures:}
Note that there is no unique choice of o-minimal structure $\cS$ of $\bbR^n$. The simplest example is the smallest structure 
that contains all the algebraic sets. It is given by collecting all semi-algebraic subsets 
of $\bbR^n$ and will be denoted by  $\bbR_{\rm alg}$. These sets can be defined by polynomials $P(x_1,...,x_n)=0$ and polynomial inequalities $P(x_1,...,x_n)>0$, in 
$n$ variables, together with their finite unions. There are, however,  various notable extensions of the smallest 
o-minimal structure $\bbR_{\rm alg}$ that are relevant in the following: 
\begin{itemize}
\item[(A):] An o-minimal structure, denoted by $\bbR_{\rm exp}$, is generated by $\bbR_{\rm alg}$ and the graph of 
the real exponential exp$: \bbR \rightarrow \bbR$ as shown in \cite{Wilkie96}. This implies that this structure is generated by 
all sets given by exponential polynomial equations $P(x_1,...,x_m,e^{x_1},...,e^{x_m})=0$ and projections thereof.
\item[(B):] An o-minimal structure, denoted by $\bbR_{\rm an}$, is generated by  $\bbR_{\rm alg}$ extended by the graphs 
of all restricted real analytic functions. Such functions are all restrictions $f|_{B(R)}$ of functions $f$
that are real analytic on a ball $B(R')$ of finite radius $R'$ to a ball $B(R)$ of strictly smaller radius $R<R'$. 
 
\item[(C):] An o-minimal structure, denoted by $\bbR_{\rm an,exp}$, is generated by $\bbR_{\rm alg}$ extended by the graphs 
of the exponential function and all restricted real analytic functions.  
\end{itemize}
Let us stress that it is a non-trivial task to find extensions of $\bbR_{\rm alg}$ that preserve o-minimality and 
Wilkie's deep theorem that $\bbR_{\rm exp}$ is o-minimal is an example of this fact. However, it will be clear  
from the concrete applications to string theory effective actions that these extensions are crucially needed. 
In fact, we will see that we will be quickly led to use the o-minimal structure $\bbR_{\rm an,exp}$ as soon 
as exponential corrections, e.g.~arising from instantons, play a role. For this reason the Tameness Conjecture
is referring to $\bbR_{\rm an,exp}$.\footnote{Note that there are several other known extensions of $\bbR_{\rm alg}$.} 

\begin{figure}[h!]
\vspace*{.4cm}
\begin{center}
 \includegraphics[width=0.6\textwidth]{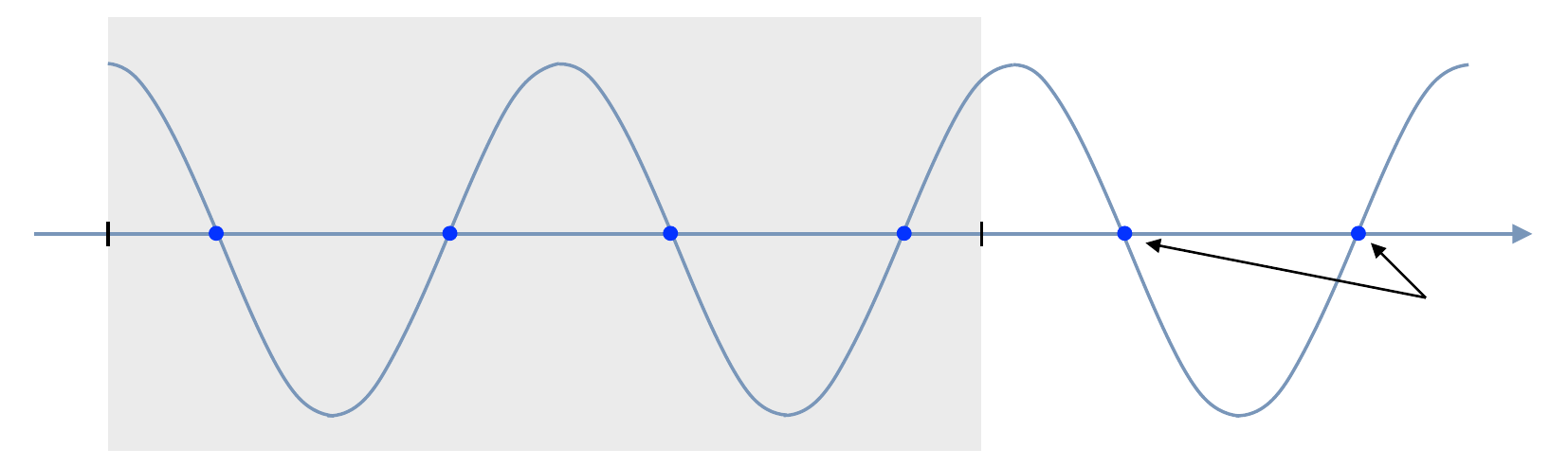} \hspace*{1.5cm}
\caption{Graph of cosine function on $\bbR$ is non-definable set, but becomes definable when restricted to domain $0\leq \phi \leq a$. \label{Cosinegraph}}
\end{center}
\begin{picture}(0,0)
\put(318,82){\small infinite, discrete zero-set}
\put(84,86){\small 0}
\put(235,87){\small $a$}
\end{picture}
\vspace*{-.6cm}
\end{figure}
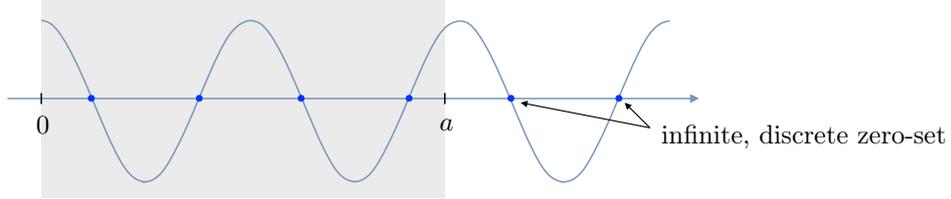

\noindent
\textbf{Complex exponential function:}
To further highlight the relevance of  $\bbR_{\rm an,exp}$, let us briefly discuss the complex exponential $e^z : \bbC \rightarrow \bbC$. Firstly, note 
that this function with domain $\bbC$ is never definable. This follows from the 
general fact that any definable and holomorphic function $f: \bbC^n \rightarrow \bbC$ has 
to be algebraic \cite{vdD}. A more direct way to see that $e^z$ is never definable on $\bbC$ is to write 
$e^{z} = e^{r + i \phi} = e^{r}(\cos \phi + i \sin \phi)$. However, the graph of the sine- and cosine-functions on 
all of $\bbR$, cannot be definable, since the projection to the $\phi$-axis gives an infinite discrete set of zeros (see figure \ref{Cosinegraph}). 
To make $e^z$ definable, we first have to restrict the domain of $z$, say by demanding $0\leq \phi \leq a$. This 
resolves the issue of periodicity since $\cos(\phi)$ is definable in $\bbR_{\rm an}$, when restricting the domain if $\phi$. 
$e^r$ is, however, not in $\bbR_{\rm an}$ and we are thus lead to consider $\bbR_{\rm an, exp}$ to have 
a definable $e^z$ on the domain $0\leq \phi \leq a$.

\noindent
\textbf{Functions not definable in $\bbR_{\rm an,exp}$:} As shown for the complex exponential, definability depends on the 
domain on which one considers a function. In the following we list a number of functions and domains, which have been shown in \cite{VANDENDRIES200161} to 
be not definable in $\bbR_{\rm an, exp}$. Firstly, we have the non-definability of the Gamma-function and Zeta-function
\beq
  \Gamma(x) = \int_0^\infty e^{-t} t^{x-1} dt\ , \qquad \zeta(x) = \sum_{n=1}^\infty \frac{1}{n^x}\ ,
\eeq
when restricting the domain to $(0,\infty)$ and $(1,\infty)$, respectively. Secondly, also the error function $ \int_0^x e^{-t^2} dt$
and the logarithmic integral $\int_x^\infty t^{-1} e^{-t} dt$ are not definable in $\bbR_{\rm an, exp}$ for $x \in \bbR$. It should be 
noted, however, that there can be cases in which an o-minimal structure exists that make these functions definable. This
has been shown for $\Gamma(x)$, $\zeta(x)$ in \cite{Dries2000TheFO}.
 
\subsection{On definable functions and the cell decomposition} \label{sec:definable-functions+cells}

In the following we want to summarize some basic results about $\cS$-definable functions and 
$\cS$-definable sets \cite{vdD}. As above we will drop the symbol $\cS$ if the statement is 
true for any o-minimal structure, but reintroduce it when making statements concerning 
a special structure. 

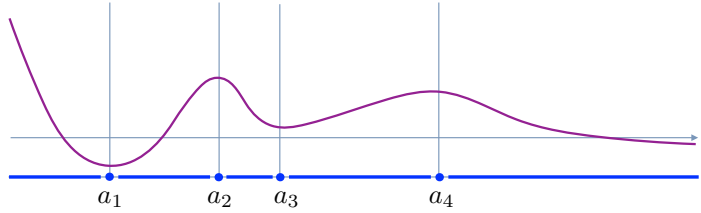
\begin{figure}[h!]
\vspace*{.4cm}
\begin{center}
 \includegraphics[width=0.6\textwidth]{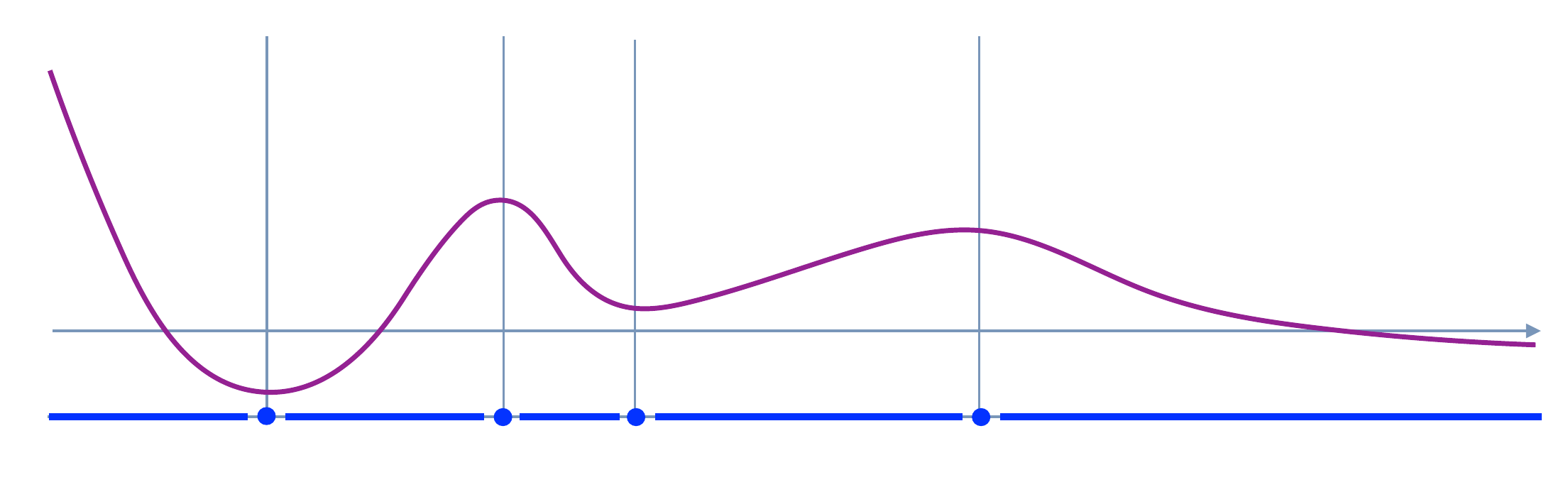}  \hspace*{.5cm}
\caption{Sketch of the graph of a definable function from $\bbR \rightarrow \bbR$ and the split of the domain into finitely many open intervals. \label{fig:Definable-function}}
\end{center}
\begin{picture}(0,0)
\put(123,62){\small $a_1$}
\put(164,62){\small $a_2$}
\put(189,62){\small $a_3$}
\put(247,62){\small $a_4$}
\end{picture}
\vspace*{-.6cm}
\end{figure}

\noindent
\textbf{Definable functions in one dimension:} Let us begin by considering a definable function $f:(a,b) \rightarrow \bbR$. The open interval $(a,b) \subset \bbR$ can 
be of finite or infinite size, including the whole $\bbR$. Definability of $f$ implies \cite{vdD} that $(a,b)$ admits a finite 
subdivision, i.e.~a split
\beq \label{interval-decomp}
   a=:a_0 < a_1 < ...< a_{m-1} < a_m:=b\ ,
\eeq
with the property that $f$ on the intervals $(a_k,a_{k+1})$ is either constant or strictly monotonic and continuous.
In particular, this implies that $f$ can only have finitely many discontinuities. One can even go one step further and 
show that there always exists a finite (possibly finer) decomposition of $(a,b)$, such that $f$ is once (or even multiple times) differentiable on the resulting 
open intervals. It also follows that $f$ can only admit finitely many minima and maxima. We depict a definable function in figure \ref{fig:Definable-function}.

Let us note that the notion of having a definable function has both local as well as global implications. To highlight one other implication let us now consider the 
o-minimal structure $\bbR_{\rm an,exp}$. If we consider a definable function in this o-minimal structure $f: \bbR \rightarrow \bbR$, we realize that for each such 
function there will be two infinitely long intervals $(-\infty,a_1)$ and $(a_{m-1},\infty)$ along which 
the function is either constant or continuous and strictly monotonic. Since $\bbR_{\rm an,exp}$ only offers restricted analytic functions, they will actually not be relevant in  
some appropriately chosen subintervals $(-\infty,\tilde a_1)$ and $(\tilde a_{m-1}, \infty)$. This implies that 
in these `asymptotic regions' of $\bbR$ the definable functions have either an algebraic or an exponential behavior. 

\noindent
\textbf{Definable cylindrical cell decomposition:} The use of the decomposition \eqref{interval-decomp} of the interval $(a,b)$ hints towards a more general strategy that 
applies to dealing with definable sets and functions. More precisely, we will now introduce a definable cylindrical cell 
decomposition of $\bbR^n$. The following discussion might, at first, look rather technical and 
can be skipped in first reading. However, eventually the resulting description 
of definable sets is the base of many subsequent theorems in the study of o-minimal structures and 
gives an intuitive understanding about the properties of higher-dimensional definable sets and functions. 
To describe a  definable cylindrical cell decomposition, we first note that it is a partition of $\bbR^n = \cup_i D_i$ into
finitely many pairwise disjoint definable subsets $D_i$, which are called cells. The crucial part is that these cells 
have special inductive description:
\begin{itemize}
 \item For $n=0$, i.e.~$\bbR^0$, there is a unique cell, which is simply all of $\bbR^0$, i.e.~a point. 
 \item For $n=1$, i.e.~$\bbR$, the cells are obtained by a decomposition \eqref{interval-decomp} of the interval $(-\infty, \infty)$. They 
 consist of the points $\{a_k \}$ for $0<k<m$, and the open intervals $(a_k,a_{k+1})$ for $0\leq k < m$. We 
 depict such a decomposition in figure \ref{def-decR}. 
 \begin{figure}[h!]
\begin{center}
\vspace*{.5cm}
 \includegraphics[width=0.6\textwidth]{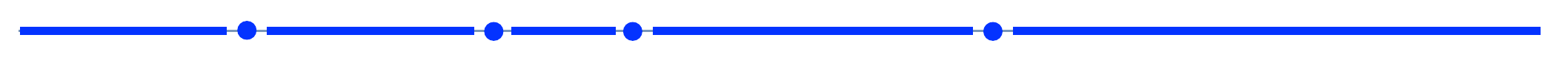} 
\caption{A definable cylindrical cell decomposition of $\bbR$. \label{def-decR}}
\end{center}
\begin{picture}(0,0)
\put(132,58){\small $a_1$}
\put(171,58){\small $a_2$}
\put(196,58){\small $a_3$}
\put(260,58){\small $a_4$}
\end{picture}
\vspace*{-.8cm}
\end{figure}
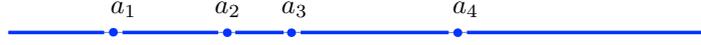
 \item In general, for $n>0$, we write $\bbR^n = \bbR^{n-1} \times \bbR$. Now we can assume that we have a definable cylindrical cell decomposition $\{ \cD_\alpha\}$ for $\bbR^{n-1}$. For each cell $\cD_\alpha$ we now have 
an integer $m_\alpha>0$ and definable continuous functions $f^{(\alpha)}_k: \cD_\alpha \rightarrow \bbR$ for $0< k < m_\alpha$ 
such that 
\beq
   -\infty =: f^{(\alpha)}_0  < f^{(\alpha)}_1 < \ldots < f^{(\alpha)}_{m_\alpha -1} < f^{(\alpha)}_{m_{\alpha}} := \infty\ ,  
\eeq
where the inequalities are meant to hold on all of $\cD_\alpha$. Having such a set of functions the cells in $\bbR^n$ are: \\
(1) graphs of the functions, i.e.~$\{(x,f^{(\alpha)}_k(x))\subset \bbR^n:x\in \cD_\alpha \}$ for each 
$\cD_\alpha$;  \\
(2) bands between the functions, i.e.~$\{(x,y)\subset \bbR^n: x\in \cD_\alpha, \text{ $y$ in  interval}\ (f^{(\alpha)}_k(x),f^{(\alpha)}_{k+1}(x))\}$. 
\end{itemize}
Due to its iterative nature, the definition of a definable cylindrical cell decomposition uses an ordering 
of the coordinates. The arising cells are thus admitting special directions along which there is a simple 
projection to a low-dimensional cell decomposition. We illustrate this in figure \ref{def-decR2}, where we depict a 
definable cylindrical cell decomposition of $\bbR^2$ build from the decomposition of $\bbR$ depicted in figure \ref{def-decR}.

 \begin{figure}[h!]
\begin{center}
\vspace*{.5cm}
 \includegraphics[width=0.65\textwidth]{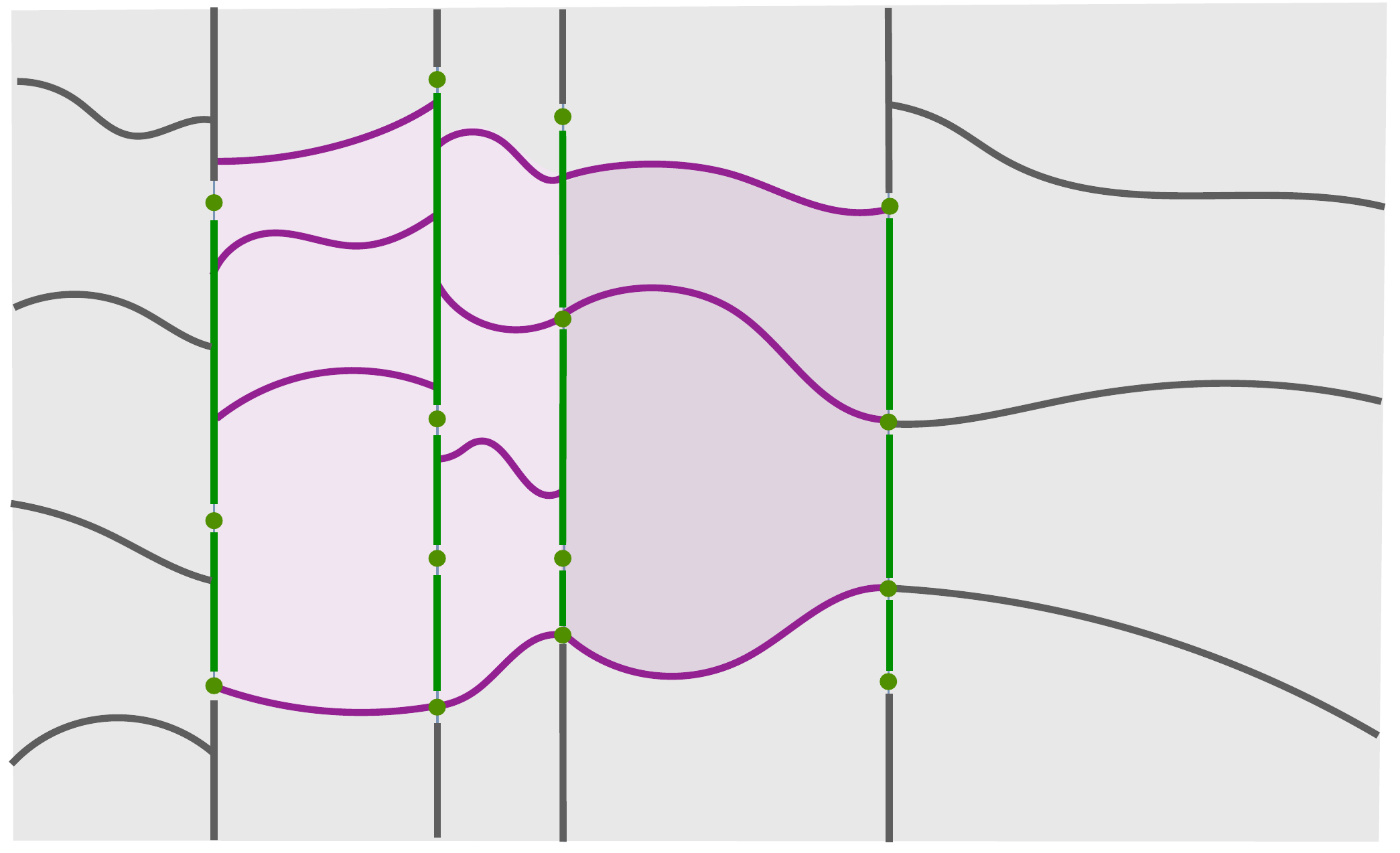} 
 \vspace*{.5cm}
\caption{A definable cylindrical cell decomposition of $\bbR^2$ build from the decomposition in figure \ref{def-decR}. The new $\bbR^2$-cells
are graphs and bands of definable functions on the $\bbR$-cells. The new cells over 
the open intervals $(a_1,a_2)$, ..., $(a_3,a_4)$ are shown in purple, while the new cells over the points $a_1,...,a_4$ are shown in green. 
We have coloured all cells stretching to $\pm \infty$ in grey.  \label{def-decR2}}
\end{center}
\begin{picture}(0,0)
\put(120,92){\small $a_1$}
\put(167,92){\small $a_2$}
\put(194,92){\small $a_3$}
\put(260,92){\small $a_4$}
\end{picture}
\vspace*{-.8cm}
\end{figure}
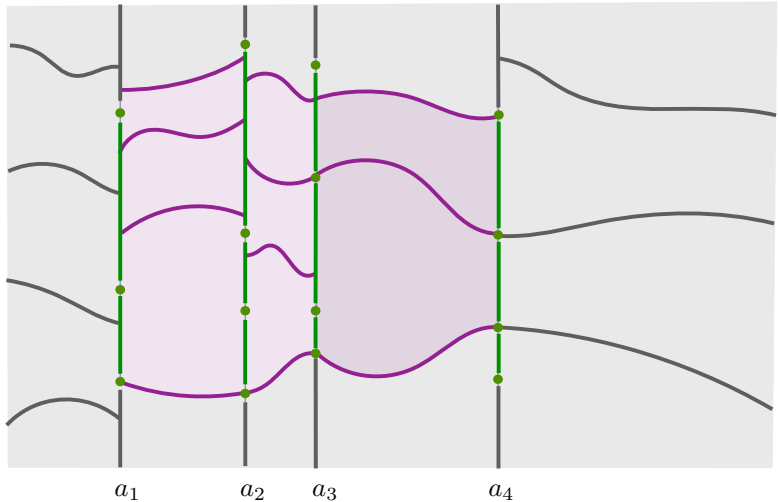

\noindent
\textbf{Cell decomposition theorem:} The remarkable fact about the cell decomposition is that they are
sufficient to describe any definable set. In fact, one can show the cell decomposition theorem: (1) Given any finite collection of definable sets $A_1,...,A_k \in \bbR^n$ there is a definable cylindrical cell decomposition such that each $A_i$ is a finite union of cells; (2)
For each definable function $f:A\rightarrow \bbR$, $A\subset \bbR^n$ there is a cylindrical cell decomposition of $\bbR^n$
such that partitions $A$ as in (1) in such a way that $f$ restricted to each cell is continuous. This theorem can be 
further refined by replacing the requirement of having continuous functions in the definition of the cells by having 
functions that are once (or multiple times) differentiable. In this case the cells might be smaller, but the cell decomposition 
theorem still holds. 

\noindent
\textbf{Dimension and Euler characteristic:}  The cell decomposition theorem has many applications. For example, it 
can be used to show 
that one can associate a dimension and an Euler characteristic to each definable set. Let us consider 
a definable set $A$ and denote by $\cD_i$ the finite number of cells in which it can be partitioned using 
the cell decomposition theorem. The  
dimension $\text{dim}(A)$ of $A$ is simply defined to be maximum found for the dimensions $ \text{dim}(\cD_i)$ of the cells $\cD_i$. 
The Euler characteristic is defined to be 
\beq
   E(A) = \sum_{\cD_i} (-1)^{ \text{dim}(\cD_i)} = k_0 - k_1 +...+(-1)^{\text{dim}(A)} k_{\text{dim}(A)} \ ,  
\eeq
where $k_s$ is the number of $s$-dimensional cells in $\{\cD_i \}$. Crucially, the values of $\text{dim}(A)$ and 
$E(A)$ are independent 
of the chosen cell decomposition and hence can serve as invariants associated to the definable set $A$.

\noindent
\textbf{Definable family and uniform boundedness:} In the concrete applications to effective theories, the notion of having a definable family of sets will be useful. To define such a family 
we consider a definable set $S \subset \bbR^{m+ n} = \bbR^m \times \bbR^n$. We then introduce the subsets $S_\lambda \subset \bbR^n$ by setting 
\beq \label{def-deffamily}
    S_{\lambda}= \{x \in \bbR^n : (\lambda,x) \in S \}\ . 
\eeq
The sets $S_\lambda$ are the fibers of the definable family $\{ S_\lambda\}_{\lambda \in \bbR^m}$.\footnote{Note that in tame topology one often calls 
$\bbR^m$ the parameter 
space of the definable family. This is in conflict with what we called parameter space $\cP$ in \eqref{def-domain} and we will reserve 
the name for $\cP$.} Note that the definable family is defined over all of $\bbR^m$, but will have empty fibers at points that do not lie in the 
projection of $S$ to $\bbR^m$. 
The cell decomposition theorem can 
now be used to prove uniform boundedness results for definable families. 
For example, consider a definable family $S_\lambda$. Then there 
exists a positive integer $m$, such that $S_\lambda$ has at most $m$ isolated points. In particular,
each fiber containing only a finite number of points has at most $m$ such points. 

Definability in an o-minimal structure has found numerous applications 
in geometry. In particular, it is interesting to point out that `definability' has been used in various 
theorems as a replacement of stronger properties of geometric spaces in algebraic geometry. 
To quote two influential theorems in which definability replaces compactness, let us 
mention the definable Chow theorem \cite{peterzil_starchenko_2008} and the Pila-Wilkie theorem \cite{PilaWilkie}. 

This ends our brief account on o-minimal structures and tame geometry. It is important to stress that this is a broad 
and well-developed field and the preceding summary should be seen more as an invitation to the field rather than 
aiming at giving a complete account.

\section{Tame geometry of the string landscape} \label{tamegeom-landscape}

In this section we discuss the evidence for the validity of the Tameness Conjecture. Before doing so, we want to use the mathematical 
background introduced in section \ref{intro-to-o} to elaborate on the 
statement of the conjecture itself. 

The Tameness Conjecture makes the assertion that the allowed parameter spaces, scalar field spaces, and coupling functions
are definable in an o-minimal structure. While at first this statement deals with very 
different objects, we now realize from subsection \ref{sec:definablesets+maps} that we should understand the parameter space $\cP$, and the field spaces $\cM_\lambda$
as subsets of $\bbR^n$ and $\bbR^m$ for sufficiently large $n$ and $m$, respectively. The coupling functions we then understand as maps from these subsets into suitable target spaces that are also embedded into some real Euclidean ambient space. 
As discussed in the beginning of 
section \ref{sec:effective_theories} the field space $\cM_\lambda$ can depend on the parameters chosen from $\cP$ and, therefore, should  be understood 
as being part of the combined set $\cD$ defined in \eqref{def-domain}, which has fibers $\cM_\lambda$. The definability statement now 
asserts that the set $\cD$, understood as a subset of $\bbR^{m+n}$, is a definable set in some o-minimal structure. We can also use the notion of definable family 
introduced in subsection \ref{sec:definable-functions+cells}, see \eqref{def-deffamily}. The Tameness Conjecture then implies the statement:  
\beq \label{definable-family}
  \text{The scalar field spaces $\{\cM_{\lambda}\}_{\lambda \in \bbR^n}$ form a definable family, where $\cM_{\lambda} = \emptyset$ if $\lambda \notin \cP$.} 
\eeq 
If $\cM_\lambda$ would be merely a set, a non-trivial consequence of definability is the fact that there 
is a well-defined $\text{dim}(\cM_\lambda) $ with $\text{dim}(\cM_\lambda) \leq n$ for all $\lambda$. 
This fits with our assertion that $\cM_{\lambda}$ has more structure, since it is considered to be the field 
space of some scalars $\phi^i$. In particular, we want to endow $\cM_{\lambda}$ with a metric to define the kinetic terms of $\phi^i$. Hence, we require $\cM_{\lambda}$ to be a Riemannian manifold. The 
definability statement then amounts to the statement that $\cM_{\lambda}$ is a definable manifold with a definable metric $g: T\cM_\lambda \times T\cM_\lambda \rightarrow \bbR$. The statement \eqref{definable-family} can then be strengthened to: $\{\cM_{\lambda}\}_{\lambda \in \bbR^n}$ forms a definable family of Riemannian manifolds. We similarly proceed for the other coupling functions in the effective theory. If  a coupling function 
admits some additional property, the Tameness Conjecture asserts that definability in an o-minimal structure should arise as an 
additional and compatible feature. 
 
In the remainder of this section we provide evidence for the Tameness Conjecture by going 
through the various compactifications mentioned 
in section~\ref{sec:effective_theories} and introduce some recent mathematical results that confirm definability in 
the o-minimal structure $\bbR_{\rm an, exp}$. In subsection \ref{sec:8ormore} we comment on string theory 
effective actions with 8 or more supercharges and exploit the fact that arithmetic quotients
are definable manifolds and that period mappings are definable maps. 
In subsection \ref{sec:N=1def} we sketch the proof that also the Type IIB/F-theory flux landscape is definable.

\subsection{Effective theories with extended supersymmetry} \label{sec:8ormore}

In order to provide evidence for the Tameness Conjecture we will first comment on the higher-supersymmetric settings 
and then turn to Calabi-Yau threefold compactifications with $\cN=2$ supersymmetry. 
The settings that we are going to discuss have been already introduced in subsection \ref{sec:Scalar-coupling}. 

\noindent 
\textbf{Definability in higher-supersymmetric settings.} In theories with more than 8 supercharges we recalled that 
the moduli spaces are arithmetic quotients \eqref{moduli-space-quotient}. Fixing the groups $G,K$ and the lattice $\Gamma$, it 
is a central result of Bakker, Klingler, Tsimerman \cite{BKT} that the manifolds $\cM_{\Gamma,G,K} = \Gamma \backslash G/K$
are definable in the o-minimal structure $\bbR_{\rm alg}$. Remarkably, the definable structure of $\cM_{\Gamma,G,K} $
is inherited from the natural definable structure of $G/K$.\footnote{Let us note that the precise statement requires 
us to introduce so-called Siegel sets, which takes much technical effort. These sets can be used to specify a 
definable atlas of $G/K$. Given a Siegel set $\mathfrak{S} \subset G/K$ one can then show \cite{BKT} that the map 
$\pi|_{\mathfrak{S}}: \mathfrak{S} \rightarrow S_{\Gamma,G,K}$ is definable in $\bbR_{\rm alg}$.} 
This implies that the field spaces for these theories are definable also in $\bbR_{\rm an,exp}$ containing $\bbR_{\rm alg}$. 
Furthermore, it should then be not too hard to check that also the coupling functions varying 
over $\cM_{\Gamma,G,K}$ are definable in $\bbR_{\rm alg}$ due to the fact that they are given by polynomial expressions. 

More subtle is the question if $\cM_{\Gamma,G,K}$ and the coupling functions are also definable when considered jointly with the parameter space $\cP$.
Note that in general there are infinitely many choices for the groups $(G,K,\Gamma)$. Each choice we consider as being labelled by a discrete 
parameter in the space $\cP$. Whether or not the allowed set is finite is, at least in some of the settings, still an open question. For example, 
consider a six-dimensional effective theory with $\cN=(1,0)$ supersymmetry. This theory has 8 supercharges, but the scalar field space in the tensor and vector 
sector of the theory is still an arithmetic quotient with $G=SO(1,T)$, where $T$ is the number of tensor multiplets. 
Bounds on $T$ and general finiteness statements about such six-dimensional theories were recently discussed in \cite{Tarazi:2021duw}. Evidence in this direction can 
therefore be directly interpreted as evidence for the Tameness Conjecture. Conversely, assuming the Tameness Conjecture a finiteness constraint on $T$
is a necessary criterion, since infinite discrete sets are never definable. 

\noindent 
\textbf{Definability in Calabi-Yau threefold compactifications with $\cN=2$.}
Let us now turn to the four-dimensional supergravity theory with $\cN=2$ supersymmetry, i.e.~8 supercharges, 
that arise when compactifying  Type IIB string theory on a Calabi-Yau threefold. We have introduced some 
basics on these settings already in subsection \ref{sec:Scalar-coupling}. Recall that supersymmetry implies that 
the field space spanned by the complex scalars in the vector multiplets is a special K\"ahler manifold $\cM_{\rm cs}$. 
The relevant local metric $g_{i\bar \jmath}$ on this manifold takes the form \eqref{N=2metric}, while the 
gauge coupling functions $f_{AB}$ for the vector fields was given in \eqref{N=2gaugecoupling}. Both can be 
expressed in terms of the periods $\Pi$ of the $(3,0)$-form $\Omega$ introduced in \eqref{periods}. 
Note that supersymmetry already implies that $g_{i\bar \jmath}$ and $f_{AB}$ can be expressed in 
terms of a holomorphic function $\Pi$, but there are no general constraints on $\Pi$ that go beyond the 
special geometry relations (see, e.g.~\cite{Craps:1997gp}, for an introduction to this subject). In Calabi-Yau compactifications 
$\Pi$ is much more constrained, since it arises from a so-called period mapping. In fact, it is a very remarkable result of 
Bakker, Klingler, and Tsimerman \cite{BKT} that the period mappings are definable in $\bbR_{\rm an,exp}$. 

To introduce the precise statement we first recall some facts about $\cM_{\rm cs}$ and then
explain the notion of a period mapping. In preparation for the discussion of the scalar potential in 
subsection \ref{sec:N=1def} we will present the following discussion for a general Calabi-Yau manifold 
of complex dimension $D$. 
For a polarized Calabi-Yau $D$-fold  the moduli space $\cM_{\rm cs}$ is complex quasi-projective \cite{Viehweg} and 
smooth after possibly performing a resolution \cite{Hironaka}. We can view $\cM_{\rm cs}$ as an 
$\bbR_{\rm an,exp}$-definable manifold by extending its $\bbR_{\rm alg}$-definable manifold structure. 
To introduce the period mapping, our starting point is the 
Hodge decomposition of the middle cohomology of $Y_D$.  Let us thus consider 
the decomposition 
\beq \label{Hodge-decomposition}
    \cH_\bbC = \bigoplus_{p+q=D} H^{p,q}\ ,
\eeq
where $\cH_\bbC$ is the primitive part of the middle cohomology $\cH_\bbC = H^{D}_{\rm prim}(Y_D,\bbC)$, i.e.~we impose $J\wedge \omega=0$ for $\omega \in \cH_\bbC$ and $J$ being a K\"ahler form on $Y_D$. 
Note that $H^3_{\rm prim}(Y_3) = H^3(Y_3)$, if we require that $H^1(Y_3)$ vanishes.  
Importantly, the decomposition \eqref{Hodge-decomposition} depends on the point in $\cM_{\rm cs}$ at which it is 
evaluated. The period map $h$, which in turn determines the period integrals, encodes this dependence by expressing 
the relation of the $H^{p,q}$ at some point $z$ with respect to a reference point $H^{p,q}_{\rm ref}$. Concretely, let us 
define $h$ as 
\beq
     H^{p,q} = h(z,\bar z) H^{p,q}_{\rm ref}\ , 
\eeq
where $h$ can be represented by a matrix acting on fixed basis of $\cH_\bbC$.
This allows us to identify the period integrals 
\beq \label{Pifromh}
   \Pi = \int_{\gamma_I} h \Omega_{\rm ref}\ , 
\eeq
where $\Omega_{\rm ref}$ is representing the one-dimensional space $H^{D,0}_{\rm ref}$. We note that $h$ becomes holomorphic when 
evaluated on a suitable basis $F^p_{\rm ref} = \bigoplus_{k=p}^D H^{k,D-k}$.

The map $h$ can be understood as maps into arithmetic quotients of the form \eqref{moduli-space-quotient}. 
To 
see this, we first introduce two real groups $V \subset G \subset \text{Gl}(\cH_\bbR)$. 
To define $G$, we first introduce  on $ \cH_\bbC$ the  
bilinear form 
\beq \label{bilin1}
   (a,b)\equiv  \int_{Y_D} a \wedge b\  , 
\eeq
where $a,b \in \cH_\bbC$.
The group $G$  consist of all elements in $\text{Gl}(\cH_\bbR)$ that preserve this bilinear form, i.e.~obey $(ga,gb)= (a,b)$. A subgroup 
of $G$, denoted by $V$,  is the group of elements that additionally preserve the whole reference $(p,q)$-splitting, 
i.e.~obey $v H^{p,q}_{\rm ref} =H^{p,q}_{\rm ref}$.
Up to global symmetries, which we will discuss in a moment, one can use these groups to identify $h$ as a map
$h : \cM_{\rm cs} \rightarrow G/V$. In order to discuss the global symmetries, note that $\cM_{\rm cs}$ is not simply connected. The monodromy group 
$\Gamma$, being a representation of the fundamental group of $\cM_{\rm cs}$ on $\cH_\bbZ=\cH_\bbC \cap H^{D}(M,\bbZ)$, captures the information about the non-trivial fibration structure of the Hodge decomposition $H^{p,q}$ arising 
due to this fact. Hence, $h$ should actually be viewed as a map 
\beq
   h :\ \cM_{\rm cs} \rightarrow \Gamma \backslash G/V. 
\eeq
A foundational result of Bakker, Klingler, Tsimerman \cite{BKT} is the theorem: 
\beq \label{tameperiodmap}
  \boxed{\rule[-.1cm]{0cm}{.5cm}\quad \text{The period mapping $h$ is definable in the o-minimal structure} \ \bbR_{\rm an,exp}\ . \quad }
\eeq
While we will not aim at reviewing the details of the proof of this statement, a few remarks might help to illuminate the steps that go into the argument. 
Firstly, as mentioned above, $h$ can be viewed as a holomorphic map, just like $\Pi$. The essential part of the proof 
is then to control the asymptotic behavior near the boundaries 
of $\cM_{\rm cs}$, since we can `discard' the compact region making up the interior of the moduli space. 
This is due to the fact that $\bbR_{\rm an,exp}$-definable functions include 
 any restricted analytic function, which endows us with a sufficiently large set of choices for this compact region.  
 The fact that the asymptotic form of $h$ is constrained as shown by Schmid \cite{Schmid}, 
see \eqref{nilp-orbit} for the analog statement for $\Pi$, suffices to establish that $h$ is compatible 
with $\bbR_{\rm an,exp}$-definability at least before modding out by $\Gamma$. To show that the quotient 
by $\Gamma$ does not ruin definability is more involved and requires to use another important result 
of asymptotic Hodge theory, namely the sl(2)-orbit theorem \cite{Schmid,CKS}. 

We are now in the position to discuss the validity of the Tameness Conjecture in these $\cN=2$ settings 
using the fact \eqref{tameperiodmap}. Since the periods $\Pi$ are given by \eqref{Pifromh}
they are also definable in $\bbR_{\rm an,exp}$. This fact can now be used in the expressions 
\eqref{N=2metric} and \eqref{N=2gaugecoupling} for the field space metric $g_{i\bar \jmath}$ and the gauge coupling 
functions $f_{\alpha \beta}$ to establish their definability in $\bbR_{\rm an,exp}$ over $\cM_{\rm cs}$. 
Hence, we have assembled another non-trivial piece of evidence for the Tameness Conjecture.
Let us stress that our analysis only establishes definability over the moduli space $\cM_{\rm cs}$. It is 
well-known that the periods $\Pi$ also depend on parameters that are fixed in terms of the geometric 
data of $Y_D$. For example, $\Pi$ near the large complex structure point in $\cM_{\rm cs}$ depends 
on the topological data of the mirror Calabi-Yau manifold associated to $Y_D$, such as its intersection numbers and Chern classes.  
The parameter space $\cP$ therefore contains a discrete set of data and definability would be lost if this set is 
infinite. In particular, it is a consequence of the Tameness Conjecture that the number of topologically distinct compact Calabi-Yau 
manifolds is finite (see \cite{Wall1966} for a more precise notion of distinguishing Calabi-Yau manifolds). 
Establishing this finiteness statement would thus be a central test of the Tameness Conjecture. 
While we will not be able to address finiteness of geometries, the 
next subsection will be devoted to establishing a non-trivial definability result over 
another discrete parameter space, namely a flux lattice.

\subsection{Definability of the flux landscape} \label{sec:N=1def}

In this subsection we discuss a definability statement that establishes
the Tameness Conjecture being satisfied over a discrete parameter space. More precisely,
we will study flux compactifications introduced in subsection \ref{sec:scalar_pot} and show that 
the scalar potential as a function of the complex structure deformations and the flux 
parameters is definable close to its self-dual vacuum locus~\eqref{G4-self-dual}. We will summarize 
the proof of this statement by following the work of Bakker, Schnell, Tsimerman, and the 
author \cite{BGST}. For simplicity, we restrict the following arguments to a study of $G_4$ flux
on a Calabi-Yau fourfold that yields a scalar potential \eqref{potential_F}.\footnote{The theorems 
proved in \cite{BGST} are much more general and hold for any variation of Hodge structures and 
thus, in particular, for any compact K\"ahler manifold.}

To begin with, we introduce in addition to \eqref{bilin1} a second bi-linear form on $\cH_{\bbC} = H^{4}_{\rm prim}(Y_4,\bbC)$
which is associated with the Hodge norm by setting
\beq
  \qquad \langle a|b \rangle \equiv  (\bar a,Cb) =  \int_{Y_D} \bar a \wedge \star b \ ,
\eeq
and we denote $\|a\|^2=\langle a|a\rangle$ as in \eqref{Hodge-positive}.
Here we have introduced the Weil operator $C$, which is nothing else than the Hodge star acting on elements 
of the cohomology. Note that $C$ acts on elements in $H^{p,q}$ with an eigenvalue $(-1)^{\frac{1}{2}(p-q)}$ and hence satisfies 
$C^2 = 1$ for even $D$ and $C^2=-1$ for odd $D$. Just as the periods $\Pi$ and the period mapping $h$, also $C$
will vary over the complex structure moduli space $\cM_{\rm cs}$. To describe this behavior we 
again fix a reference Hodge decomposition $H^{p,q}_{\rm ref}$ and an associated 
Weil operator $C_{\rm ref}$. 
The Weil operator  at the point $z$ in $\cM_{\rm cs}$ can be obtained from $C_{\rm ref}$ 
by using the period mapping introduced in the previous section by 
\beq \label{C-hrelation}
 C(z,\bar z) = h^{-1} C_{\rm ref} h
\eeq 
but it turns out to be better to consider $C(z,\bar z)$ directly and study its properties as a map from 
$\cM_{\rm cs}$ into some quotient space. To find this quotient we note that every Weil operator $C'$
can be obtained from $C_{\rm ref}$ by acting with an element $g \in G$ as $C' = g C_{\rm ref} g^{-1}$. For later use, let us denote this 
operator by  
\beq \label{def-Cg}
   C_g := g C_{\rm ref} g^{-1}\ .
\eeq
Denoting by $K$ the group elements preserving $C_{\rm ref}$ we thus identify $C$ as a map 
\beq
   C:  \cM_{\rm cs} \rightarrow  G/K\ .
\eeq 
Here the symmetric space $G/K$ labels all inequivalent Weil operators that can be defined on $\cH_\bbC$. 

\noindent
\textbf{Scalar potential for fixed flux.} Let us now turn to the analysis of the flux scalar potential. As 
a first step, we will fix the flux $G_4$ and only consider the dependence on $\cM_{\rm cs}$. 
In this case the argument for $V$ being definable in $\bbR_{\rm an, exp}$ is analog to the analysis of 
the field space metric and gauge coupling function outlined in subsection \ref{sec:8ormore}. 
Recall that the Hodge star in \eqref{potential_F} reduces 
to $C$ on cohomology classes, and hence we can write in the notation of this section 
\beq \label{V(z,v)}
   V(z,v) = c \| C v - v \|^2 \ .
\eeq
As we move along $\cM_{\rm cs}$ the Hodge decomposition will vary and hence also 
the associated Weil operator. It now follows from \eqref{C-hrelation} and the definability of the period map 
\eqref{tameperiodmap} that 
 \beq \label{C-definable}
  \boxed{\rule[-.1cm]{0cm}{.45cm}\quad \text{The Weil operator $C$ is definable in the o-minimal structure} \ \bbR_{\rm an,exp}\ . \quad }
\eeq
Since $\| \cdot \|$ is built using $C$, we readily apply \eqref{C-definable} to conclude that
\beq \label{V-definable}
   V(G_4): \ \cM_{\rm cs} \rightarrow \bbR\ \ \text{is}\ \ \bbR_{\rm an,exp}\text{-definable}\ .  
\eeq
By a generalization of the statements of subsection \ref{sec:definable-functions+cells} one sees that it therefore has only finitely many disconnected sets of zeros and minima. Note that 
this is also true if we replace $V$ with any definable function of $C$.

It is important to stress at this point that the statement \eqref{V-definable} only holds when fixing the flux $G_4$. Since, $G_4$
takes values on a lattice, definability as a function of the parameter $G_4$ will be lost if no further constraints 
are imposed on $G_4$. It is not hard 
to see that also the tadpole constraint $(G_4,G_4)=\ell$
still allows for infinitely many choices of $G_4$ and hence does not suffice to ensure definability. In the next step we will see, however, that 
$V$ as a function of $G_4$ is actually definable near self-dual vacua when imposing the tadpole constraint.

 \noindent
\textbf{Definability and self-dual fluxes.} Let us now also take into account that one can choose the fluxes $G_4$ in the 
scalar potential from a lattice $\cH_\bbZ = H^{4}(Y_4,\bbZ) \cap H^{4}_{\rm prim}(Y_4,\bbR)$ as long as they satisfy the tadpole constraint. 
We sketch the proof that definability is
retained in the product of $\cM_{\rm cs}$ and the flux lattice 
when considering the zeros of $V(v,z)$ given in \eqref{V(z,v)}. More precisely, let us introduce the Hodge bundle $E$ with fibers $\cH_\bbC$, 
which encodes the variation of the $(p,q)$-decomposition of $\cH_\bbC$ when moving over the base $\cM_{\rm cs}$. 
Note that $E$ is an algebraic bundle and hence is a definable manifold in $\bbR_{\rm alg}\subset \bbR_{\rm an,exp}$. 
Our aim is to study the subsets of $E$ at which the integral fluxes in the fibers of $E$ satisfy the self-duality and 
the tadpole condition. The statement proved in \cite{BGST} is 
\beq \label{self-flux-def}
 \boxed{\rule[-.4cm]{0cm}{1.1cm}\quad \begin{array}{l} \text{The subset $\{ (z,G_4) \subset E$ with fluxes $G_4$ s.t.~$C(z) G_4 = G_4,(G_{4} , G_4)=\ell \}$ }\\[.1cm]
 \text{is definable in the o-minimal structure} \ \bbR_{\rm an,exp}\ . \end{array} \quad }
\eeq
In particular, this includes the observation that a reduction of $E$ to this set has finite fibers. Using the statements 
about definable families and uniform boundedness from subsection \ref{sec:definable-functions+cells} one thus concludes that 
there are only finitely many fluxes $G_4$ that possibly can satisfy the self-duality and tadpole conditions. 

To elucidate some of the steps that go into showing \eqref{self-flux-def}, let us fix a $\cH_\bbZ$ not changing over $\cM_{\rm cs}$ 
and note the all integral elements in $E$ can be reached from this $\cH_{\bbZ}$ up to monodromy. 
Hence we want to study the sets 
\beq   \label{subsets_bundle}
    (\cM_\alpha ,G^\alpha_4) \subset \cM_{\rm cs} \times \cH_\bbZ \ , \qquad {\alpha = 1,2,...} \ ,
\eeq
 by requiring 
 \bea
    (1) &&(G_4^\alpha,G_4^\alpha) = \ell \ , \\
    (2) &&C(z,\bar z) G_4^\alpha = G_4^\alpha\quad  \forall z \in \cM_\alpha \ ,
 \eea
 where no sum over $\alpha$ needs to be performed in $(1)$.
To obtain these sets, we pick a flux $G^\alpha$ satisfying (1) and then determine all points in $\cM_{\rm cs}$ that are obeying (2). 
At first, since there infinitely many choices for $G^\alpha_4 \in \cH_\bbZ$, the index range of $\alpha$ is infinite. Furthermore, 
also $\cM_\alpha$ could have infinitely many disconnected components. For the second statement, however, definability of $C$ 
as stated in \eqref{C-definable} actually ensures that $\cM_\alpha$ is $\bbR_{\rm an,exp}$-definable, and hence has only finitely many 
connected components. We now want to show that the index range of $\alpha$ is actually finite. 
 
Let us introduce the symmetry group $\Gamma_\cH$ of the lattice $\cH_\bbZ$ preserving the inner product by setting 
 \beq
 \Gamma_{\cH} =  O(\cH_\bbZ, (\cdot,\cdot) )\ , 
\eeq
An important step in \cite{BGST} is to use this symmetry group and reduce the lattice into finitely many orbits of $\Gamma_\cH$ 
along which we then are able to show definability. To do that we use a result 
of Kneser \cite{Kneser:QuadratischeFormen} on lattices and bilinear forms. Let us pick a $G_4\in \cH_\bbZ$ and act with all elements in $\Gamma_\cH$ on 
$G_4$ to define the equivalence class $[G_4]$. Kneser now shows that the set of fluxes $G_4$ with a fixed $(G_4,G_4)=\ell$ is obtained from 
only \textit{finitely many} such classes. In other words, one can select finitely many fluxes 
\beq \label{finite-setG4}
    G_4^A\ ,  \quad A=1,...,n \ :  \qquad (G^A_4,G^A_4)=\ell \ ,
\eeq 
and generate the whole 
set of solutions to $(G_4,G_4)=\ell$ by acting with $\Gamma_\cH$. Remarkably, the 
tadpole condition thus gives us a reduction to checking definability in finitely many orbits $\Gamma_\cH G_4^A$.

Let us fix a reference Weil operator $C_{\rm ref}$ as in \eqref{C-hrelation} and 
pick one flux $F\in \cH_\bbZ$ that obeys $(F,F)=\ell$ and is self-dual with respect to $C_{\rm ref}$, $C_{\rm ref} F =F$. 
Clearly, $F$ can be taken to be one of the $G_4^A$ in \eqref{finite-setG4} from which we generate a $\Gamma_\cH$-orbit. 
We now want to consider all $C_g$ introduced in \eqref{def-Cg} that preserve self-duality of $F$, i.e.~we will look at the set 
\beq \label{F-set}
 \big\{ (gK ,F):\ C_g F = F \big\}_{g\in G} \ \subset\  G/K \times \cH_\bbZ\ ,
\eeq 
where we recall that each set $gK$ represents a Weil operator via \eqref{def-Cg}, since $K$ preserves $C_{\rm ref}$.
Looking at sets \eqref{F-set} is analogous 
to \eqref{subsets_bundle}, but we now work with sets representing Weil operators instead of subsets of $\cM_{\rm cs}$. 
It will be the final key step to ensure that going from $\cM_{\rm cs}$ to Weil operators can be done in 
an $\bbR_{\rm an,exp}$-definable way. We note that the equations $C_g F = F$ and $(F,F)=\ell$ have 
the symmetry 
\beq
     C_g \rightarrow k C_g k^{-1}\ , \qquad F \rightarrow k F\ , \qquad k \in \Gamma_\cH\ . 
\eeq
Hence, it will suffice the think about the set $\Gamma_\cH (gK ,F)$, where the action of the group $\Gamma_\cH$ is via $\alpha (gK, F) = (\alpha gK, \alpha F) $, 
and work on the quotient  
\beq
    \Gamma_\cH \backslash (G/K \times \cH_\bbZ)\ .
\eeq


 Let us now consider the orbit $\Gamma_\cH F \subset \cH_\bbZ$ generated when acting with all elements of $\Gamma_\cH$ on $F$. 
 To begin with, we define the real groups $G_F, K_F, \Gamma_{F}$ 
as the subgroups preserving $F$, i.e.~we set
\beq
   G_F =\{ h\in G : h F= F\}\ , \qquad K_F = K \cap G_F\ , \qquad \Gamma_F = \Gamma_\cH \cap G_F\ . 
\eeq
Since $F$ is self-dual with respect to $C_{\rm ref}$ it is an element in $G_F/K_F$. In fact, the symmetric space $G_F/K_F$ labels all 
Weil operators $C_g$ that fix the element $F$. 
We now consider a $F_\gamma = \gamma F \in \Gamma_\bbZ F$ that is also self-dual with 
respect to $C_g$. This implies that we can write
\beq
   F = \gamma^{-1} F_\gamma = \gamma^{-1} C_g F_\gamma = \gamma^{-1} C_g \gamma F = C_{ \gamma^{-1} g } F \ .
\eeq
Since $C_{ \gamma^{-1} g }$ is a Weil operator fixing $F$ there should be an $h \in G_F$ such that $C_h = C_{ \gamma^{-1} g }$.
Reading this as the condition  $C_{\rm ref} = C_{h^{-1} \gamma^{-1} g}$ we conclude that $h^{-1} \gamma^{-1} g \in K$. This implies that 
\beq
   \Gamma_\cH ( g K, F_\gamma ) =   \Gamma_\cH ( h K, F ) \ .
\eeq
This relation implies that the set of $\{ F_\gamma\}_{\gamma \in \Gamma_\cH}$ with $C_g F_\gamma = F_\gamma$ is actually the image of a map  
\beq
   i : \Gamma_F \backslash G_F/K_F \rightarrow \Gamma_\cH \backslash (G/K \times \cH_\bbZ)\ , \qquad  \Gamma_F h K_F \mapsto \Gamma_\cH (h K ,F)\ . 
\eeq
However, by of another result of \cite{BKT} (see also \cite{BGST}), such maps between algebraic quotients are actually $\bbR_{\rm alg}$-definable. 
The locus of Weil operators mod $\Gamma_\cH$ and self-dual classes in $\Gamma_\cH F$, i.e. 
\beq \label{subsets1}
 \big\{\ \Gamma_{\cH}(g K , F_\gamma) : \ C_g F_\gamma = F_\gamma \ \big\}_{g \in G,\gamma \in \Gamma_\cH} \ \subset\  \Gamma_\cH \backslash (G/K \times \cH_\bbZ)
\eeq
is therefore an $\bbR_{\rm alg}$-definable subset that is isomorphic to the smaller
arithmetic quotient $\Gamma_F \backslash G_F/K_F $.

It remains to show that definability of the set \eqref{subsets1} in $ \Gamma_\cH \backslash (G/K \times \cH_\bbZ)$ can be carried over to 
the space $\cM_{\rm cs} \times \cH_\bbZ$. This actually follows from an extension of the definability property of the Weil operator \eqref{C-definable}. 
In fact, in order to show the definability of the period mapping \eqref{tameperiodmap}, the authors of \cite{BKT} actually first prove the 
definability of $C$ mod $\Gamma_\cH$. Let us define the Weil operator period map 
\beq \label{C-map1}
   [C]: \ \cM_{\rm cs} \rightarrow \Gamma_\cH \backslash G/K\ , 
\eeq
which associates to each point in $\cM_{\rm cs}$ its Weil operator modulo $\Gamma_\cH$. The definability statement then reads \cite{BKT}
\beq \label{C-definable}
  \boxed{\rule[-.1cm]{0cm}{.45cm}\quad \text{The Weil operator period map $[C]$ is definable in the o-minimal structure} \ \bbR_{\rm an,exp}\ . \quad }
\eeq
We stress that this is a stronger statement than the definability of the period mapping stated in \eqref{tameperiodmap}, since the latter involves 
 the monodromy group $\Gamma$ and $\Gamma \subset \Gamma_\cH$. Finally, one has to extend the map \eqref{C-map1} to 
 an $\bbR_{\rm an,exp}$-definable map $E \rightarrow  \Gamma_\cH \backslash (G/K \times \cH_\bbZ)$. 
 This is straightforward if one thinks about $E$ being the product $\cM_{\rm cs} \times \cH_\bbC $, but requires some 
 extra work to incorporate the bundle structure in a definable way as explained in \cite{BGST,flatbun}. Since the pre-image of the 
 sets \eqref{subsets1} are precisely the self-dual integral classes satisfying the tadpole constraint, and a definable 
 set under a definable map is definable, we can conclude the statement \eqref{self-flux-def}.
 
Let us close this section by stressing that \eqref{self-flux-def} is a statement about the global minima
of $V(z,G_4)$, which does not imply definability for every minimum of $V(z,G_4)$ when 
we allow changes of $z$ and $G_4$. Whether or not a more general statement about 
all minima of $V(z,G_4)$ can be proved is an open question. On the one hand, 
one can try to extend the approach of \cite{BGST}, maybe restricting attention solely to the 
Calabi-Yau fourfold case discussed here. On the other hand, it can very well be the case that such a more 
general statement is simply not true. This would indicate that there exist 
infinitely many vacua with broken supersymmetry due to non-vanishing F-terms for the complex structure moduli (see 
\cite{Denef:2004cf} for a study of such settings). If one would be able to trust all the effective theories arising 
near these vacua this would be a clear violation of the Tameness Conjecture. While we cannot make any 
conclusive statements on this, let us note that the string embedding of the flux backgrounds that are 
not self-dual is more obscure and one might argue that these vacua simply do not yield controllable effective theories.     
In contrast, recalling \eqref{V-definable}, we can consider $V(z,G_4)$ near its self-dual vacua $V(z,G_4)=0$ and conclude that 
the Tameness Conjecture is satisfied for $V$ as a function of over the accessible field space and the parameter space of 
allowed fluxes.

\section{Conclusions and discussions}
\label{sec:con}
In this work we have proposed a Tameness Conjecture, which states all effective theories compatible 
with quantum gravity are labelled by a definable parameter space and must have scalar field spaces and coupling functions 
that are definable in an o-minimal structure. Here one considers the set of all effective theories valid below a fixed 
finite cut-off scale. 
The weak version of 
this conjecture asserts that any o-minimal structure can be used, while the stronger 
version fixes the underlying o-minimal structure to be $\bbR_{\rm an,exp}$. 
This choice of o-minimal structure was supported by all examples of string theory effective 
actions. Independent of the precise choice of o-minimal structure, the resulting 
tame geometry has strong finiteness properties and thus imposes structural constraints on 
attainable parameter spaces, field spaces, and coupling functions. Accordingly, our initial 
motivation for these condition is the conjectured finiteness of the set of effective theories 
arising from string theory \cite{Douglas:2003um,Acharya:2006zw,Hamada:2021yxy}. The Tameness Conjecture implements this in an intriguing 
way. On the one hand, the definability of the parameter space imposes 
that there are only finitely many `disconnected' choices to obtain a scalar field space and coupling 
functions. On the other hand, definability of the scalar field space and coupling functions then 
ensures that an initial effective theory admits only finitely many effective theories when lowering the 
cut-off. While finiteness was the central motivation, tame geometry actually provides us with a set of 
 local and global constraints that go beyond finiteness restrictions that we expect are relevant to  
further connect some of the swampland conjectures. 

To provide evidence for the Tameness Conjecture we have analyzed various effective theories 
that arise after compactifying string theory. While for ten-dimensional supergravity theories the 
conjecture is readily checked at the level of the two-derivative effective action, it becomes increasingly hard to 
test it in full generality when going to lower dimensions. This can be traced back to the facts 
that (1) supersymmetry does not necessarily strongly constrain the form of the field space and the coupling functions, 
and (2) there is an increasing number of parameters in the theory. For more than 
8 supercharges, one still finds that definability of the field spaces with fixed parameters follows from supersymmetry. 
We have argued that this is due to the fact that they are given as  
arithmetic quotients $\Gamma \backslash G / K$ that are definable in $\bbR_{\rm alg}$ for a fixed 
choice of $G,\Gamma,K$. String theory then has to ensure that there are only finitely many choices 
of parameters, e.g.~only finitely many allowed groups $G,\Gamma,K$.
For settings with $8$ or less supercharges also field spaces with fixed parameters can be non-definable, 
since supersymmetry is not strong enough to ensure the presence of the tameness properties. 
We have shown, however, that in string compactifications, in particular on Calabi-Yau manifolds, 
the  non-trivial constraints of the allowed deformations of the compactification geometry ensure definability. 
More concretely, we have seen that the complex structure and K\"ahler structure moduli space of 
Calabi-Yau manifolds are definable 
and admit a physical metric that is definable. This latter fact is a consequence of the non-trivial fact that 
the period mapping is definable in $\bbR_{\rm an, exp}$ as recently shown in \cite{BKT}. By using the 
definability of the period mapping we also concluded the definability of the gauge coupling functions in four-dimensional 
 Type II compactifications on Calabi-Yau threefolds.

As the most involved test of the Tameness Conjecture we studied Type IIB and F-theory flux compactifications 
yielding to a four-dimensional theory with $\cN=1$ supersymmetry. A non-trivial background flux 
induces a scalar potential and we investigated in detail its tameness properties. We have found 
that for fixed fluxes, this scalar potential is definable as a consequence of the definability of the 
Hodge star operator. When allowing to also change the flux, definability appears to be lost, since 
the flux takes values on an infinite discrete set even after imposing the tadpole constraint. However, 
we have shown that definability is restored when constraining the attention to effective theories 
near self-dual flux vacua. To see this, we have sketched the proof of \cite{BGST} that the locus 
of self-dual flux vacua is definable in $\bbR_{\rm an,exp}$ even if one collects all possible flux 
choices consistent with the tadpole constraint. In other words, the Tameness Conjecture for the 
scalar potential is even satisfied over the discrete parameter space set by the fluxes, if we take into 
account the required existence of a self-dual flux vacuum. We have discussed that the latter 
constraint might be necessary since only in the self-dual cases one has $V(z_{\rm vac})=0$ 
and there is a clean higher-dimensional 
description of the vacuum in Type IIB or F-theory. These facts might be needed to justify the 
notion of working in a well-defined effective theory. 
Alternatively, if one aims to extend this result to other vacua of $V$ one might have to impose additional 
conditions on $V(z_{\rm vac})$ or the masses of the scalars to retain definability for a theory at fixed cut-off. 
This example already highlights many of the issues 
that arise in any theory with a scalar potential. In particular, we assert that the 
Tameness Conjecture remains satisfied when lowering the cut-off and integrating 
out fields. Our results show that for self-dual flux vacua one can send the cut-off 
to zero and obtain a new effective theory with only massless complex structure moduli 
that is definable in the considered sector. 

Let us now turn to a more general discussion of the statement and the implications 
of the Tameness Conjecture. We will collect some thoughts on our findings and indicate 
some future directions for research:

\noindent
\textbf{Tameness and gravity.} The Tameness Conjecture has been formulated as a requirement on effective theories 
that can be consistently coupled to quantum gravity. However, from its formulation it 
is not apparent which role gravity plays in its statements. From the study of examples we 
note that gravity genuinely appears to constrain the parameter space of the effective theory. The 
Tameness Conjecture states that the parameter space should never include infinite discrete sets. However, without 
considering a UV-completion it is not hard to find infinite sets of supersymmetric theories with field spaces that are individually 
definable but have no bound on the parameter labelling the dimension of these spaces. It is believed that gravity will eventually provide us with 
a bound on the maximal dimensionality of allowed field spaces and hence restrict the associated discrete parameter 
space to a definable set. We have seen something analogous happening in our flux compactification example, 
where the tadpole constraint, which is a crucial consistency condition on 
compact internal manifolds, was needed as a key element to reduce to finitely many flux orbits. 
It remains to provide more tests of the reduction to finite discrete sets when it comes 
to geometry. The Tameness Conjecture implies, in particular, that there should be only finitely many 
topologically distinct manifolds that one can choose to obtain valid effective theories.
In Calabi-Yau compactifications this seems to require the finiteness of topologically 
distinct compact Calabi-Yau manifolds. Moreover, validity of the effective theory 
can impose constraints on curvatures and volumes of the compactification space,
and it has been discussed in \cite{Acharya:2006zw} that these can 
lead to a reduction to finitely many topological types by a theorem of Cheeger \cite{Cheeger}.
While these arguments support definability of the parameter space, it would be interesting 
to provide a more in-depth study of the necessary minimal conditions on the compactification spaces.

\noindent 
\textbf{Tameness and other swampland conjectures.} It is an interesting open question to investigate connections 
between the Tameness Conjecture and other well-known swampland conjectures beyond the ones mentioned above. 
Note that tame geometry is a rather flexible framework, which allowed us to suggest that any effective theory, in particular 
also without supersymmetry, can be covered by the Tameness Conjecture. Hence, due to the novel nature of the 
constraints imposed by the Tameness Conjecture, we would not expect that 
it directly implies any of the other conjectures. In fact, one may expect that this conjecture 
becomes really powerful when combined with additional constraints that have been suggested before. In particular, 
the Tameness Conjecture suggests some interesting interrelation with the Distance Conjecture \cite{Ooguri:2006in} and the Emergence 
Proposal \cite{Harlow:2015lma,Heidenreich:2017sim,Grimm:2018ohb,Heidenreich:2018kpg,Corvilain:2018lgw,Palti:2019pca}. We have explained that every definable function has a more 
constrained `tame' behavior in non-compact directions. It would be interesting to see if this fact can be 
linked with the Distance Conjecture when considering infinite distance directions in field space. Furthermore, it 
might be that the existence of tame non-compact directions in field space is only an emergent phenomenon
that arises when integrating out states of an underlying quantum gravity theory. If this were true it would imply that 
the Tameness Conjecture actually imposes general constraints on the degrees of freedom and their interaction in the 
underlying fundamental theory. It is an 
exciting task to test this idea for simple examples and we hope to return to this in a future work.        

\noindent
\textbf{Tameness replacing compactness.} Let us point out that the Tameness Conjecture for field spaces also offers a more general 
perspective on the properties of brane moduli space. While for lower-dimensional branes 
these moduli spaces were conjectured to be compact \cite{Hamada:2021bbz,Bedroya:2021fbu}, it is well-known that for higher-dimensional 
branes, such as 7-branes in Type IIB string theory compactness is not a suitable criterion. However, 
it is known from the geometric realization of 7-branes in an elliptically fibered Calabi-Yau manifold in 
F-theory that the moduli space of these extended objects is definable in an o-minimal structure. In other 
words, while colliding 7-branes can admit a non-compact moduli space, the geometry of this space 
is tame in the asymptotic direction. It would thus be interesting to investigate whether one finds direct arguments 
for the Tameness Conjecture by analyzing the physics of 7-branes in non-compact directions. 
Conversely, we have mentioned already that tame geometry provides strong theorems 
that replace compactness with tameness and we expect that they can be used to prove general results about  
the behavior of 7-branes in F-theory.

\noindent
\textbf{Tameness and the classification of effective theories.} Another remarkable implication of the Tameness Conjecture is that 
it allows for a novel way to classify effective theories. The triangulation theorem in tame geometry states that any 
definable set is definably homeomorphic to a polyhedron \cite{vdD}. This identification occurs if and only if the sets and the polyhedron have 
the same dimension and Euler characteristic introduced in subsection \ref{sec:definable-functions+cells}. The triangulation theorem states that the 
topological information in the set can be described in finite combinatorial terms. Hence, it provides a new way to compare the information defining 
two effective theories by comparing their parameter spaces, field spaces, and coupling functions as definable sets. It would 
be very interesting to explore this for simple quantum field theories or conformal field theories. The definability in an o-minimal structure 
hereby can serve as an additional structure on the space of theories that could allow to further extend the ideas put forward in \cite{Douglas:2010ic}.

\subsubsection*{Acknowledgments}

I am indebted to Benjamin Bakker, Christian Schnell, and Jacob Tsimerman for their collaboration on \cite{BGST}, which motivated this work. 
In particular, I am grateful to Christian Schnell for  
introducing me to the subject and answering many of my questions. 
I would also like to thank Michael Douglas, Stefano Lanza, Chongchuo Li, Jeroen Monnee, Miguel Montero, Eran Palti, 
Damian van de Heisteeg, Cumrun Vafa, Stefan Vandoren, 
and Mick van Vliet for insightful discussions and correspondence.
This research 
is partly supported by the Dutch Research Council (NWO) via a Start-Up grant and a Vici grant.

\appendix

\bibliographystyle{jhep}
\bibliography{Mini_Bib}

\end{document}